\documentclass[camera,letterpaper,nomarginnotes,nonarrowgutter]{jpaper}

\usepackage{dblfloatfix}
\usepackage{multirow}
\usepackage{booktabs}
\usepackage{array}
\usepackage{algorithm}
\usepackage{eucal}
\usepackage{rotating}
\usepackage{xcolor,colortbl}
\definecolor{freakishgreen}{HTML}{0A982B}
\definecolor{urlblue}{HTML}{319dd6}
\usepackage{ifthen} %
\usepackage{xspace} %
\usepackage{enumitem}%
\usepackage{fancyhdr}
\usepackage{enumitem}%
\usepackage[noend]{algpseudocode}
\usepackage{datetime}
\usepackage{textcomp} %
\usepackage{xparse} %
\usepackage{setspace} %
\usepackage{minted} %
\newcolumntype{L}[1]{>{\raggedright\let\newline\\\arraybackslash\hspace{0pt}}m{#1}}
\newcolumntype{C}[1]{>{\centering\let\newline\\\arraybackslash\hspace{0pt}}m{#1}}
\newcolumntype{R}[1]{>{\raggedleft\let\newline\\\arraybackslash\hspace{0pt}}m{#1}}
\newcolumntype{M}[1]{>{\centering\arraybackslash}m{#1}}

\usepackage{tikz}

\makeatletter
\def\thickhline{%
  \noalign{\ifnum0=`}\fi\hrule \@height \thickarrayrulewidth \futurelet
   \reserved@a\@xthickhline}
\def\@xthickhline{\ifx\reserved@a\thickhline
               \vskip\doublerulesep
               \vskip-\thickarrayrulewidth
             \fi
      \ifnum0=`{\fi}}
\makeatother
\newlength{\thickarrayrulewidth}
\usepackage{pifont}

\def\thickhline{\noalign{\hrule height.8pt}}
\newcolumntype{?}{!{\vrule width 0.8pt}}

\usepackage{xhfill}

\usepackage{soul}
\setul{0.5ex}{0.3ex}
\setulcolor{blue}

\newif\ifsubmission
\submissiontrue

\newif\ifarxiv
\arxivtrue

\definecolor{colorV1}{rgb}{0.0, 0.0, 1.0}    %
\definecolor{colorV2}{rgb}{1.0, 0.0, 1.0}    %
\definecolor{colorV3}{rgb}{0.0, 0.5, 0.0}    %
\definecolor{colorV4}{rgb}{0.6, 0.4, 0.2}    %

\newcounter{CurrentDraftVersion}

\ExplSyntaxOn

\cs_new:Npn \revcol:n #1
  {
    \int_case:nnF {#1}
      {
        {1}{blue}
        {2}{magenta}
        {3}{orange}
        {4}{cyan}
        {5}{violet}
        {6}{teal}
        {7}{purple}
        {8}{brown}
        {9}{olive}
        {10}{red}
      }
      {blue}
  }

\NewDocumentCommand{\edit}{ O{\value{CurrentDraftVersion}} m }
  {
    \int_compare:nNnTF { \value{CurrentDraftVersion} } = {1000}
      { #2 }
      {
        \int_compare:nNnTF { \value{CurrentDraftVersion} } = {100}
          { \textcolor{blue}{#2} }
          {
            \int_compare:nNnTF { #1 } = { \int_eval:n { \value{CurrentDraftVersion} - 1 } }
              { \textcolor{\revcol:n {#1}}{#2} }
              { #2 }
          }
      }
  }

\ExplSyntaxOff

\ifsubmission
    \fancypagestyle{firstpage}
    {
        \fancyhead{}

        \pagenumbering{arabic}
        \fancyfoot[C]{\large\thepage}
    }
    
\else
    
    \fancypagestyle{firstpage}
    {
        \fancyhead[C]{\textcolor{blue}{\emph{Version \arabic{CurrentDraftVersion}~---~\today, \xxivtime \ UTC}}}
        \fancyfoot[C]{\thepage}
    }
\fi

\usepackage{hyperref}

\hypersetup{
    bookmarks=true,
    breaklinks=true,
    colorlinks=true,
    linkcolor=blue,
    citecolor=blue,
    urlcolor=urlblue,
    pdfpagelabels=false,
 }

\usepackage[nameinlink]{cleveref}
\crefformat{section}{\S#2#1#3} %
\crefformat{subsection}{\S#2#1#3}
\crefformat{subsubsection}{\S#2#1#3}

\usepackage{amsmath, amssymb, amsfonts}
\usepackage{graphicx}
\usepackage{textcomp}
\usepackage{url}
\usepackage{algorithm}
\usepackage{algpseudocode}
\usepackage{multirow}
\usepackage{tikz}
\usepackage{marginnote}
\usepackage{fancyhdr}
\usepackage{titlesec}
\usepackage{dblfloatfix}
\usepackage{makecell}
\usepackage{comment}
\usepackage{xspace}
\usepackage{xcolor, soul}
\usepackage{enumitem}
\usepackage{tabularx} %
\PassOptionsToPackage{bookmarks=true,breaklinks=true,letterpaper=true,colorlinks,citecolor=blue,linkcolor=blue,urlcolor=blue}{hyperref}
\usepackage{hyperref}
\usepackage{setspace}
\usepackage[colorinlistoftodos,prependcaption,textsize=scriptsize]{todonotes} %
\usepackage[sort,compress]{cite}

\usepackage{siunitx}
\usepackage{datetime}

\definecolor{bleudefrance}{rgb}{0.19, 0.55, 0.91}
\definecolor{caribbeangreen}{rgb}{0.0, 0.8, 0.6}
\definecolor{english}{rgb}{0.0, 0.5, 0.0}
\definecolor{teal}{rgb}{0.0, 0.5, 0.5}
\definecolor{tyrianpurple}{rgb}{0.4, 0.01, 0.24}
\definecolor{violet}{rgb}{0.56, 0.0, 1.0}
\definecolor{darkorange}{rgb}{0.78, 0.43, 0}

\newcommand{\ak}[1]{{\color{black}#1}}

\newcommand{\rncami}[1]{{\color{black}#1}}
\newcommand{\rncamii}[1]{{\color{black}#1}}
\newcommand{\rncamiii}[1]{{\color{black}#1}}
\newcommand{\rncamiv}[1]{{\color{black}#1}}
\newcommand{\rncamv}[1]{{\color{black}#1}}
\newcommand{\rncamvi}[1]{{\color{black}#1}}
\newcommand{\rncamvii}[1]{{\color{black}#1}}

\newcommand{\figs}[1]{{Figs.#1}} %
\newcommand{\fig}[1]{{Fig.#1}} %

\newcommand{\head}[1]{{\noindent\textbf{#1.}}} %

\newcommand*\circled[1]{\tikz[baseline=(char.base)]{
            \node[shape=circle,draw,inner sep=0.4pt,fill=black, text=white] (char) {#1};}}

\newcommand*\whitecircle[1]{\tikz[baseline=(char.base)]{
            \node[shape=circle,draw,inner sep=0.4pt, text=black] (char) {#1};}}

\newcommand*\circleds[1]{\tikz[baseline=(char.base)]{
            \node[shape=circle,draw,inner sep=-0.2pt,fill=black, text=white] (char) {#1};}}

\newcommand{\fc}{{Flash-Cosmos}}
\newcommand{\af}{{Ares-Flash}}
\newcommand{\isp}{{ISP}}
\newcommand{\pud}{{PuD-SSD}}
\newcommand{\bw}{{BW-Offloading}}
\newcommand{\dm}{{DM-Offloading}}
\newcommand{\ideal}{{Ideal}}
\newcommand{\cpu}{{CPU}}
\newcommand{\ifp}{{IFP}}
\newcommand{\namePaper}{{Conduit}} %

\newcommand{\squishlist}{
 \begin{list}{$\circ$}
  { \setlength{\itemsep}{0pt}
     \setlength{\parsep}{0pt}
     \setlength{\topsep}{3pt}
     \setlength{\partopsep}{0pt}
     \setlength{\leftmargin}{1em}
     \setlength{\labelwidth}{1em}
     \setlength{\labelsep}{0.5em} } }

\newcommand{\squishend}{  \end{list}  }

\def\BibTeX{{\rm B\kern-.05em{\sc i\kern-.025em b}\kern-.08em
    T\kern-.1667em\lower.7ex\hbox{E}\kern-.125emX}}

\title{\namePaper{}: Programmer-Transparent Near-Data Processing \\Using Multiple Compute-Capable Resources in \rncami{Solid State Drives}}

\author{
    Rakesh Nadig\textsuperscript{$\dagger$} \hspace{0.5em} Vamanan Arulchelvan\textsuperscript{$\dagger$} \hspace{0.5em} Mayank Kabra\textsuperscript{$\dagger$} \hspace{0.5em} Harshita Gupta\textsuperscript{$\dagger$} \vspace{0.2em} \\
    Rahul Bera\textsuperscript{$\dagger$} \hspace{0.5em} Nika Mansouri Ghiasi\textsuperscript{$\dagger$} \hspace{0.5em} Nanditha Rao\textsuperscript{$\dagger$} \hspace{0.5em} Qingcai Jiang\textsuperscript{$\dagger$} \vspace{0.2em} \\
    Andreas Kosmas Kakolyris\textsuperscript{$\dagger$} \hspace{0.5em} Yu Liang\textsuperscript{$\dagger$}\textsuperscript{$\ddagger$} \hspace{0.5em} Mohammad Sadrosadati\textsuperscript{$\dagger$} \hspace{0.5em} Onur Mutlu\textsuperscript{$\dagger$} \vspace{0.5em} \\
    \textsuperscript{$\dagger$}\emph{ETH Zürich} \hspace{0.5em} \textsuperscript{$\ddagger$}\emph{Inria, Paris}
}

\begin{document}

\maketitle
\thispagestyle{firstpage}

\begin{abstract}
Near-data processing (NDP) mitigates the data movement bottleneck in modern computing systems by performing computation \rncami{close} to where the data resides.
Solid-state drives (SSDs) are well suited for NDP because they: 
(1) store large application datasets that exceed main memory capacity, and (2) contain multiple heterogeneous computation resources, \rncami{e.g.,} general-purpose embedded cores in the SSD controller, DRAM \rncami{chips}, and NAND flash chips, which enable three NDP paradigms: in-storage processing (\isp{}), processing using DRAM in the SSD (\pud{}), and in-flash processing (\ifp{}).
These resources offer massive internal parallelism and enable in-place computation, which reduces \rncami{data movement across} the memory hierarchy.

A large body of prior SSD-based NDP techniques operate in isolation, mapping computations to only one or two NDP paradigms (i.e., \isp{}, \pud{}, or \ifp{}) within the SSD.
These techniques (1) are tailored to specific workloads or kernels, (2) do not offload computations across all three NDP paradigms in the SSD and thus fail to exploit the full computational potential of an SSD, and (3) lack programmer-transparency, often requiring significant manual effort to identify offloadable code regions and map them to the SSD computation resources, which limits their general applicability and ease of deployment.
While several prior works propose techniques to partition computation between the host and near-memory accelerators, adapting these techniques to SSDs offers limited benefits because they (1) ignore the heterogeneity of the SSD computation resources, and (2) make offloading decisions based on limited factors such as \rncami{bandwidth utilization, data movement cost\rncamii{,} or memory intensity}, while ignoring key factors such as resource utilization.

We propose \namePaper{}, a general-purpose, programmer-transparent NDP framework for SSDs that accelerates a broad range of workloads by leveraging available SSD computation resources.
\namePaper{} operates in two stages. 
\rncami{At} compile time, \namePaper{} executes a custom \rncami{compiler (e.g., LLVM)}  pass that (i) vectorizes suitable application code segments into single-instruction multiple-data (SIMD) operations that align with the SSD's page layout, and (ii) embeds metadata (e.g., operation type, operand sizes) into the vectorized instructions to guide runtime offloading decisions.
\rncami{At} runtime, within the SSD, \namePaper{} performs instruction-granularity offloading by evaluating six key application and system features \rncami{(e.g., operation type, computation resource utilization, data \rncamiv{dependence delay})}, and uses a cost function to select the most suitable SSD computation resource to execute each vectorized instruction.  
We evaluate \namePaper{} and two prior NDP offloading techniques using an in-house event-driven SSD simulator on six data-intensive applications \rncami{(e.g., large language model inference and training, \rncamii{encryption})}. 
\namePaper{} outperforms the best-performing prior offloading policy by 1.8$\times$ and reduces energy consumption by 46\%, with \rncami{small} latency and storage overheads, and no additional hardware \rncami{cost}.
\end{abstract}

\section{Introduction \label{sec:introduction}}
Near-data processing (NDP) techniques \rncamii{(e.g., \cite{boroumand2018google, mutlu2019enabling, mutlu.imw13,kanev.isca15, mutlu2019processing,mutlu2022modern, wang.micro2016,mckee2004reflections, mutlu.superfri15, park2022flash, hajinazar2021simdram, seshadri-micro-2017, gu-isca-2016, gao2021parabit, barbalace_blockndp_2020, augusta2015jafar, boroumand2019conda, fernandez2020natsa, singh2019napel, gao2016hrl, lee-ieeecal-2020, singh2021fpga, medal2019, liang-fpl-2019, ghiasi2022genstore, oliveira2024mimdram, nider2020processing, hsieh.isca16, ghiasi2022alp,park2024attacc,chen2022offload,li2018cisc, li2023optimizing,maity2025unguided, olivier2019hexo,wei2022pimprof, weiner2022tmo, yang2023lambda, lincoln-hpca, jang2025inf, pan2024instattention, jaliminchecs, kang2024isp})} alleviate the performance and energy overheads caused by \emph{data movement} in modern computing systems by enabling computation close to where data resides\rncami{~\cite{mutlu2022modern, mutlu2025memory, mutlu2019processing}}. 
In \rncami{processor-centric} computing systems, the performance and energy efficiency of modern data-intensive applications 
\rncamii{(e.g., databases~\cite{doty1980magnetic, elmasri2007fundamentals, kocberber2013meet, augusta2015jafar, Oracle, idreos2012monetdb, wu2014q100, chan1998bitmap, oneil-ideas-2007, li-vldb-2014, li-sigmod-2013, goodwin-SIGIR-2017, seshadri-micro-2013, seshadri-micro-2017, seshadri-ieeecal-2015, hajinazar2021simdram, fastbit, wu-icde-1998, guz-ndp-2014, redis-bitmaps, kabra2025ciphermatch}, 
web search~\cite{ccetin2016blind, pugsley2014ndc, cho-wondp-2013, janapa2010web, ayers2018memory, zhu2013high, reddi2011mobile}, 
data analytics~\cite{jia2017understanding, jun2015bluedbm, perach-arxiv-2022, seshadri-micro-2017, torabzadehkashi-pdp-2019, lee-ieeecal-2020, jun-isca-2018, choi2015energy, panda2014data, jun2014scalable, wang2017xpro, tiwari-fast-2013, boboila-msst-2012, gao.pact15}, 
genomics~\cite{alser-bioinformatics-2017, loving-bioinformatics-2014, xin-bioinformatics-2015, cali-micro-2020, kim-genomics-2018, lander-nature-2001, altschul-jmb-1990, myers-jacm-1999, ghiasi_megis_2024, ghiasi2022genstore, papageorgiou2018genomic, soysal2025mars, alkan.naturegenetics09}, 
graph processing~\cite{shim2022gp3d, zhang2018graphp, ahn2015scalable, zhuo2019graphq, huang2020heterogeneous, song2018graphr, dai2018graphh, beamer-SC-2012, besta2021sisa, li-dac-2016, gao2021parabit, hajinazar2021simdram}, 
\rncamii{machine learning~\cite{ankit2019puma, low.vldb12, wang2020survey, sheng2023flexgen,pan2024instattention,jang2025inf, boroumand2021google, he2025papi, gu2025pim, park2024attacc, lincoln-hpca, lee2025aif, kim2023optimstore, LSTM, reinforce1992, gomezluna2022isvlsi, qlearning_ML_1992, wang2024beacongnn, yu2024cambricon, pan2024instinfer}}, 
cryptography~\cite{tuyls-springer-2005, manavski-spcom-2007, nejatollahi2020cryptopim, han-spie-1999, lee2005architecture, hajiabadi2025cassandra, lemay2021cryptographic, devadas2022designing, wang2022eager, diffie2022new, chang2011workload}, mobile workloads~\cite{boroumand2018google, wang2013architectural, liang2025ariadne, olivier2019hexo, halpern2016mobile, gupta2024relief, mobile-tensorflow, reddi2011mobile})}
are limited by data movement between the compute units (e.g., CPU, GPU) and the memory \rncami{system}.
\rncami{Many} of these workloads (e.g., graph processing, sorting, sparse matrices) perform simple computations (e.g., bitwise operations, \rncami{comparisons}) with low arithmetic intensity, making data movement the dominant cost during execution.
NDP reduces unnecessary \rncami{data movement} and enables efficient use of memory and storage bandwidth\rncami{~\cite{mutlu2025memory, mutlu2022modern, mutlu2019processing, mutlu2019enabling, mutlu2024memory,ghose2019processing}}.

A solid-state drive (SSD) (e.g.,~\cite{samsung-980pro, adatasu630,intelqlc,intels4510, intelp4610, nadig2023venice, cai-procieee-2017, micheloni-insidenand-2010, micheloni2013inside, inteloptane, samsungmlc, samsung2017znand}) is well-suited for NDP \rncami{due to at least} three reasons.
First, it stores large application datasets that often exceed main memory capacity. 
Second, it enables three heterogeneous NDP paradigms: \rncami{1)} processing using general-purpose embedded cores in the SSD controller (in-storage processing (\isp{}))~\cite{park2016storage, kabra2025ciphermatch, mailthody-micro-2019, kim-fast-2021, kang-msst-2013, torabzadehkashi-pdp-2019, seshadri-osdi-2014, wang-eurosys-2019,acharya-asplos-1998, keeton-sigmod-1998, koo-micro-2017, tiwari-fast-2013, tiwari-hotpower-2012, boboila-msst-2012, bae-cikm-2013,torabzadehkashi-ipdpsw-2018, pei-tos-2019, do-sigmod-2013, kim-infosci-2016, riedel-computer-2001,riedel-vldb-1998, liang-atc-2019,cho-wondp-2013, jun2015bluedbm, lee-ieeecal-2020, ajdari-hpca-2019,liang-fpl-2019,jun-hpec-2016, kang-tc-2021, kim-sigops-2020, lee2022smartsage, li2023ecssd, ruan2019insider, wang2016ssd1, jeong-tpds-2019, mao2012cache, gouk2024dockerssd, ghiasi_megis_2024, ghiasi2022genstore, kang-micro-2021, yavits2021giraf, kim2023optimstore, lim-icce-2021, narasimhamurthy2019sage, jun-isca-2018,  fakhry2023review, gu-isca-2016, yang2023lambda, jo2016yoursql, chen2025reis, li-atc-2021, wang2016ssd, mahapatra2025rag, pan2024instattention, wang2024beacongnn, yu2024cambricon, pan2024instinfer}, 
\rncami{2)} processing using DRAM in the SSD (\pud{})\rncami{~\cite{seshadri-micro-2017, hajinazar2021simdram, oliveira2024mimdram, besta2021sisa,seshadri-arxiv-2019, li-micro-2017, seshadri-micro-2013, seshadri-arxiv-2016-pum, deng-dac-2018,xin-hpca-2020, gao-micro-2019, yuksel2025dram, olgun2023dram,gao2022fracdram, yuksel2024functionally, seshadri-ieeecal-2015, yuksel2024simultaneous, de2025proteus}}, and 
\rncami{3)} in-flash processing (\ifp{})\rncami{~\cite{park2022flash, gao2021parabit, chen2024aresflash, kim2025crossbit,wong2024tcam, wong2025anvil, chen2024search, choi2020flash, chun2022pif, lee2025aif, wang2018three,  kabra2025ciphermatch, kang-tc-2021}} (see \S\ref{sec:background} for details). These in-situ computation capabilities reduce unnecessary data movement between the host \rncami{processor} and the SSD, and ease the burden on the memory hierarchy.
Third, \rncami{an SSD} provides high internal parallelism for concurrent data access and computation.

While a large \rncami{number} of prior works (e.g., \cite{park2016storage, kabra2025ciphermatch, mailthody-micro-2019, kim-fast-2021, kang-msst-2013, torabzadehkashi-pdp-2019, seshadri-osdi-2014, wang-eurosys-2019,acharya-asplos-1998,keeton-sigmod-1998, wang2016ssd, koo-micro-2017, tiwari-fast-2013, tiwari-hotpower-2012, boboila-msst-2012, bae-cikm-2013,torabzadehkashi-ipdpsw-2018, pei-tos-2019, do-sigmod-2013, kim-infosci-2016, riedel-computer-2001,riedel-vldb-1998, liang-atc-2019,cho-wondp-2013, jun2015bluedbm, lee-ieeecal-2020, ajdari-hpca-2019,liang-fpl-2019,jun-hpec-2016, kang-tc-2021, kim-sigops-2020, lee2022smartsage, li2023ecssd, ruan2019insider, wang2016ssd1, li-atc-2021, jeong-tpds-2019, mao2012cache, gouk2024dockerssd, ghiasi_megis_2024, ghiasi2022genstore, kang-micro-2021, yavits2021giraf, kim2023optimstore, lim-icce-2021, narasimhamurthy2019sage, jun-isca-2018, fakhry2023review, gu-isca-2016, yang2023lambda, jo2016yoursql, chen2025reis, park2022flash, gao2021parabit, chen2024aresflash, soysal2025mars, lee2017extrav}) \rncami{propose} SSD-based NDP techniques, they have two key limitations.
First, these techniques operate largely in isolation, offloading parts of the application to \emph{only} one or two SSD computation resources, which prevents them from exploiting the SSD's full computational potential.
For example, Active Flash~\cite{tiwari-fast-2013} offloads data analytics kernels to SSD controller cores (ISP) and
Flash-Cosmos~\cite{park2022flash} exploits only flash chips to accelerate bulk bitwise operations via in-flash processing.  
Second, these techniques are typically application-specific and \rncami{\emph{not}} programmer-transparent, which \rncami{limits their general applicability}. For example, MARS~\cite{soysal2025mars} accelerates raw signal genome analysis in the SSD by adding specialized hardware units in the SSD DRAM and the SSD controller, but it relies on custom data layouts and the explicit identification of offloadable sections by the programmer.
\rncami{The need for such programmer} intervention limits the generality of these techniques and their \rncami{ease of} adoption in modern storage \rncamii{and system} stacks.

\noindent \head{Limitations of Prior Offloading Approaches} 
To our knowledge, \emph{no} prior work exploits \rncamii{\emph{multiple}} SSD computation resources in a \rncamii{\emph{general-purpose}} and \rncamii{\emph{application-transparent}} manner. 
\rncami{Several} prior NDP offloading techniques (e.g.,\ak{\cite{ghiasi2022alp,park2024attacc, chen2022offload, hsieh.isca16, li2018cisc, li2023optimizing, maity2025unguided, olivier2019hexo, wei2022pimprof, weiner2022tmo, wolski2008using, yang2023lambda, hadidi2017cairo, wu2020tuning, alsop2024pim, kim2017toward, jiang20243}}) propose partitioning and mapping applications for execution between \rncami{specifically} the host and NDP units near main memory (e.g., 3D stacked memory with \rncami{general-purpose} cores in its logic layer).
\rncami{Unfortunately,} adapting these techniques to SSD-based NDP provides limited benefits because they (1) do \rncami{\emph{not}} account for the architectural heterogeneity of the SSD computation resources, which vary in their parallelism, access granularities, and computation capabilities, and (2) optimize for \emph{only} a limited set of system-level metrics such as bandwidth utilization (e.g.,~\cite{hsieh.isca16}) or data movement cost (e.g.,~\cite{ghiasi2022alp}), while ignoring key factors such as computation resource utilization\rncami{, \rncamii{operand location,} and data dependencies}.

We study the effectiveness of two prior offloading models, \bw{} \rncami{(e.g.,~\cite{wolski2008using, yang2023lambda, hsieh.isca16, hadidi2017cairo, wu2020tuning, alsop2024pim})} and \dm{} \rncami{(e.g.,~\cite{ghiasi2022alp, kim2017toward, wei2022pimprof,jiang20243})}, when applied to offloading computations within an SSD. 
Our motivational study shows that the best-performing prior model, \dm{}, shows an average performance gap of 2.5$\times$ compared to an \ideal{} offloading approach that assumes no resource contention and always selects \rncamii{the} resource with the lowest computation latency (see \S\ref{subsec:motivation_effectiveness} for more \rncami{detail}). 
These results show the need for an offloading mechanism that fully exploits the heterogeneity of SSD computation resources and makes workload- and system-aware offloading decisions. 

\rncami{\textbf{Our goal} is to enable programmer-transparent near-data processing in SSDs that (1) \rncamii{schedules and} coordinates computation across \rncamii{multiple} heterogeneous SSD computation resources and (2) makes offloading decisions that are aware of both workload characteristics and dynamic system conditions, and (3) improves the performance and energy efficiency of a wide range of applications.
To this end, we propose \textit{\textbf{\namePaper{}}}, a general-purpose programmer-transparent NDP framework that dynamically offloads fine-grained computations (at instruction granularity) to SSD controller cores, SSD DRAM chips, and flash chips.}

\namePaper{} consists of two key steps. 
First, \namePaper{} performs compile-time vectorization to identify offloadable code regions (e.g., loops with computations) and transforms them into SIMD operations that match the internal bit-level parallelism of an SSD.
Second, at runtime, \namePaper{} (i) determines the most suitable SSD computation resource to execute each vector operation using a holistic cost function, which is based on six key factors: operation type, operand location, data dependencies, resource utilization, data movement costs\rncami{,} and computation latencies, (ii) translates each vector operation to the native instruction set architecture (ISA) of the chosen SSD computation resource, and (iii) dispatches the transformed instruction to the chosen resource's execution queue. 

We evaluate \namePaper{} and six prior NDP
techniques\rncami{~\cite{arm-cortexR8, oliveira2024mimdram, park2022flash, chen2024aresflash, yang2023lambda, ghiasi2022alp}} using an event-driven SSD simulator.\footnote{We develop \namePaper{}'s simulator due to the lack of SSD simulators that support NDP. Our \rncami{simulator} inherits its core SSD model from the state-of-the-art SSD simulator, MQSim~\cite{tavakkol2018mqsim, mqsim-github}. Our NDP extensions (i.e., models for SSD computation resources) are calibrated using real-device characterization studies (e.g.,~\cite{park2022flash, chen2024aresflash}) and open-source \rncamii{infrastructures} (e.g.,~\cite{luo2023ramulator2, oliveira2024mimdram, qemu}) from prior works (See \S\ref{subsec:methodology_modeling}).} 
Our evaluation includes six data-intensive applications (see \S\ref{sec:methodology}) that cover diverse \rncami{computation} and access patterns.
\namePaper{} outperforms the best-performing prior offloading policy\rncami{~\cite{ghiasi2022alp}} by 1.8$\times$ and reduces energy consumption by 46\% on average.

This work makes the following key contributions:
\begin{itemize}[leftmargin=*]
\item We demonstrate that prior NDP offloading approaches offer limited benefits when adapted to SSDs because they (1) do not account for the architectural heterogeneity of SSDs, and (2) optimize for only a limited set of system-level metrics (e.g., data movement or bandwidth). 
\item We propose \namePaper{}, the first general-purpose programmer-transparent NDP framework that enables fine-grained \rncamii{(i.e., instruction-granularity)} offloading across \rncami{multiple} heterogeneous SSD computation resources. 
\namePaper{} dynamically determines the most suitable computation resource for each instruction using a holistic cost function that is based on multiple application and system characteristics.
\item We evaluate \namePaper{} and \rncami{six} prior offloading techniques on six real-world data-intensive applications. \namePaper{} provides significant performance and energy benefits over prior offloading techniques across all workloads.
\end{itemize}

\section{Background}
\label{sec:background}
\rncami{We} provide an overview of \rncamii{a modern} SSD architecture and its computation resources. We describe the principles of the three NDP paradigms enabled by SSDs: in-storage processing, processing using DRAM (\pud{}), and in-flash processing. 

\subsection{SSD Overview}
\fig~\ref{fig:ssd_architecture} shows a NAND flash-based SSD, which consists of three main components: the SSD controller~\circled{1}, DRAM~\circled{2}, and NAND flash chips~\circled{3}. 
The SSD controller includes multiple general-purpose embedded cores~\circled{4} (e.g.,~\cite{arm-cortexR5,arm-cortexR8}) that execute the Flash Translation Layer (FTL) firmware~\cite{gupta2009dftl, tavakkol2018mqsim, lim2010faster, shin2009ftl, zhou2015efficient} and handle \rncami{six} key functions: 
(1) communicate with the host \rncami{using} protocols such as SATA~\cite{sata} or NVMe~\cite{nvme} over system I/O bus (e.g., PCIe\rncami{~\cite{pcie-2017}})~\circled{5}, 
(2) translate the logical address of every I/O request to a physical address (L2P mapping), 
(3) schedule accesses to flash chips and DRAM via multiple chip~\circled{6} and bank-level queues~\circled{7} respectively.
(4) perform garbage collection~\cite{yang2014garbage, cai-procieee-2017, tavakkol2018mqsim, agrawal2008design, shahidi2016exploring, lee2013preemptible,jung2012taking, choi2018parallelizing,wu2016gcar,cai-insidessd-2018,cai-hpca-2017} to reclaim old and invalid blocks, 
(5) distribute writes (i.e., wear-leveling~\cite{cai-date-2012,cai-dsn-2015,cai-hpca-2015,cai-hpca-2017,cai-iccd-2012}) across flash blocks to uniformly wear out the SSD and improve the endurance, and  
(6) maintain FTL metadata and \rncami{cache} frequently accessed pages in \rncami{SSD's DRAM~\circled{2}}.

\begin{figure}[t]
    \centering
    \includegraphics[width=0.8\linewidth]{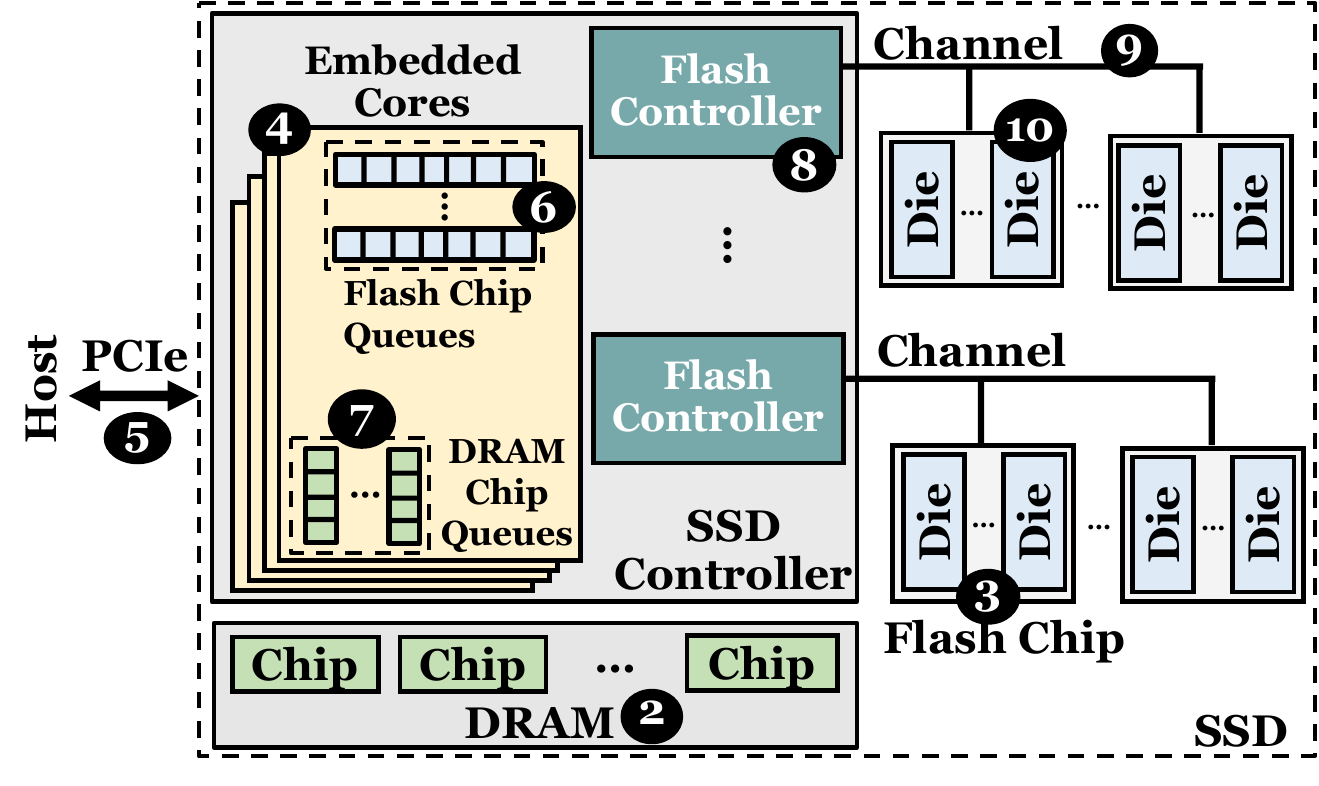}
    \caption{High-level overview of a modern SSD}
    \label{fig:ssd_architecture}
\end{figure}

A flash controller (FC)~\circled{8}\rncamii{~\cite{kim2021decoupled, kim2022networked,wu2012reducing, micheloni-insidenand-2010, microchip-16-channel-controller, cai-date-2012, cai-iccd-2012, cai-inteltechj-2013, cai-date-2013, cai-iccd-2013, cai-sigmetrics-2014, cai-hpca-2015, cai-dsn-2015, cai-hpca-2017, luo-sigmetrics-2018, cai-procieee-2017, cai-insidessd-2018, luo2018heatwatch, luo2016enabling, luo2015warm}} is a \rncami{specialized processor} that connects to flash chips\rncami{~\cite{agrawal2008design, cai-date-2012, cai-iccd-2012, cai-inteltechj-2013, cai-date-2013, cai-iccd-2013, cai-sigmetrics-2014, cai-hpca-2015, cai-dsn-2015, cai-hpca-2017, luo-sigmetrics-2018, cai-procieee-2017, cai-insidessd-2018}} via shared channels~\circled{9}. 
A modern SSD typically includes one FC per flash channel to maximize internal SSD parallelism. 
Each FC is responsible for (1) command/data transfer~\cite{micheloni-insidenand-2010,onfi-2022}, (2) data randomization~\cite{cai-procieee-2017,cai-hpca-2017,cai-insidessd-2018}, and (3) \rncami{ECC encoding/decoding}\rncamii{~\cite{cai-procieee-2017, cai-hpca-2017, cai-insidessd-2018, park2022flash,zhao2013ldpc, micheloni-insidenand-2010,tanakamaru-jssc-2013,park-asplos-2021, luo2018improving, luo2018heatwatch, luo2016enabling, luo2015warm}}. 

Each flash chip contains 1–4 independently operating dies~\circled{10}, which provide fine-grained parallelism by enabling independent access and command execution. We provide a more detailed description of the flash chip structure, and how it enables in-flash computation in \S\ref{subsec:background_ndp}.

\subsection{Near-Data Processing \label{subsec:background_ndp}}
Near-data processing (NDP)~\cite{boroumand2018google, mutlu2019enabling, mutlu.imw13,kanev.isca15, mutlu2019processing,mutlu2022modern, wang.micro2016,mckee2004reflections, mutlu.superfri15, park2022flash, hajinazar2021simdram, seshadri-micro-2017, gu-isca-2016, gao2021parabit, barbalace_blockndp_2020, augusta2015jafar, boroumand2019conda, fernandez2020natsa, singh2019napel, gao2016hrl, lee-ieeecal-2020, singh2021fpga, medal2019, liang-fpl-2019, ghiasi2022genstore, oliveira2024mimdram, nider2020processing, hsieh.isca16, ghiasi2022alp,park2024attacc,lincoln-hpca, jang2025inf, pan2024instattention, chen2022offload, jaliminchecs, kang2024isp, li2018cisc, li2023optimizing,maity2025unguided, olivier2019hexo,wei2022pimprof, weiner2022tmo, yang2023lambda, chen2024aresflash, kabra2025ciphermatch, kim2025crossbit,kang-tc-2021, chen2024search, wong2024tcam, besta2021sisa, seshadri-arxiv-2019, li-micro-2017, seshadri-micro-2013, seshadri-arxiv-2016-pum, deng-dac-2018,xin-hpca-2020, gao-micro-2019, seshadri-ieeecal-2015, yuksel2024functionally, yuksel2025dram, yuksel2024simultaneous, park2016storage, mailthody-micro-2019, kim-fast-2021, kang-msst-2013, torabzadehkashi-pdp-2019, seshadri-osdi-2014, wang-eurosys-2019,acharya-asplos-1998,keeton-sigmod-1998, wang2016ssd, koo-micro-2017, tiwari-fast-2013, tiwari-hotpower-2012, boboila-msst-2012, bae-cikm-2013,torabzadehkashi-ipdpsw-2018, pei-tos-2019, do-sigmod-2013, kim-infosci-2016, riedel-computer-2001,riedel-vldb-1998, liang-atc-2019,cho-wondp-2013, jun2015bluedbm, ajdari-hpca-2019, jun-hpec-2016, kim-sigops-2020, lee2022smartsage, li2023ecssd, ruan2019insider, wang2016ssd1, li-atc-2021, jeong-tpds-2019, mao2012cache, gouk2024dockerssd, ghiasi_megis_2024, kang-micro-2021, yavits2021giraf, kim2023optimstore, lim-icce-2021, narasimhamurthy2019sage, jun-isca-2018, fakhry2023review, jo2016yoursql, chen2025reis, mahapatra2025rag, torabzadehkashi2018compstor, olgun2023dram, gao2022fracdram,wong2025anvil, perach-arxiv-2022, gokhale1995processing, patterson-ieeemicro-1997, oskin1998active, kang1999flexram} mitigates the data movement bottleneck by enabling computation closer to the data, typically \rncami{inside} or near memory or storage \rncami{arrays}. 
SSDs are well-suited for NDP because they inherently include three heterogeneous computation resources: (1) SSD controller cores, which support general-purpose computation \rncami{capability}, (2) \rncamii{SSD-internal} DRAM, which enables fast, parallel bulk bitwise and arithmetic operations, and (3) NAND flash chips, which \rncami{enable parallel} in-place bulk bitwise and arithmetic operations. 

\head{In-Storage Processing (\isp{})}
The SSD controller cores\rncami{~\cite{arm-cortexR5,arm-cortexR8,marvellsc5bravera, armssdcontroller, cai-procieee-2017, micheloni2013inside, cheong2018flash, hatanaka2012nand, liao2012multi, park2006high, do2019improving, dirik2009performance, takeuchi2009novel,kim2021decoupled, kim2022networked}} typically execute FTL firmware and handle host communication, but they can be repurposed to perform general-purpose computations. 
Prior works (e.g.,~\cite{park2016storage, kabra2025ciphermatch, mailthody-micro-2019, kim-fast-2021, kang-msst-2013, torabzadehkashi-pdp-2019, wang-eurosys-2019,acharya-asplos-1998,keeton-sigmod-1998, tiwari-fast-2013, tiwari-hotpower-2012, boboila-msst-2012, bae-cikm-2013,torabzadehkashi-ipdpsw-2018, pei-tos-2019, do-sigmod-2013, kim-infosci-2016, riedel-computer-2001,riedel-vldb-1998, liang-atc-2019,cho-wondp-2013, jun2015bluedbm, lee-ieeecal-2020, ajdari-hpca-2019,liang-fpl-2019,jun-hpec-2016, kim-sigops-2020, lee2022smartsage, li2023ecssd, ruan2019insider, wang2016ssd1, li-atc-2021, jeong-tpds-2019, mao2012cache, gouk2024dockerssd, ghiasi_megis_2024, ghiasi2022genstore, kang-micro-2021, yavits2021giraf, kim2023optimstore, lim-icce-2021, narasimhamurthy2019sage, jun-isca-2018, fakhry2023review, gu-isca-2016, yang2023lambda, jo2016yoursql, chen2025reis, wang2016ssd, seshadri-osdi-2014, mahapatra2025rag, pan2024instattention, lee2017extrav, wang2024beacongnn, yu2024cambricon, pan2024instinfer}) demonstrate that these embedded cores can accelerate \rncami{various} tasks \rncami{including filtering, aggregation, encryption, compression and search}. 
\rncami{\rncamii{Limited} SIMD parallelism (e.g., 32-bit registers in ARM Cortex-R5~\cite{arm-cortexR5}) in the SSD controller cores reduces the computation throughput, making \isp{} less effective for data-parallel workloads.}

\head{Processing-using-DRAM in SSD (\pud{})} 
Modern SSDs integrate a \rncami{set of} low-power DRAM \rncami{chips} (e.g., 4GB LPDDR4 for 4TB SSDs~\cite{samsung-980pro}) to store metadata and cache frequently accessed pages\rncami{~\cite{gupta2009dftl, tavakkol2018mqsim, lim2010faster, shin2009ftl, zhou2015efficient, cai-procieee-2017, lim_faster_2010,sun_leaftl_2023, wang2024learnedftl, ma_lazyftl_2011, zhou_efficient_2015}}.
This DRAM subsystem exposes high internal bandwidth and parallelism, making it a well-suited substrate for NDP.
\fig~\ref{fig:processing_using_ssd_dram} shows DRAM organization and the architectural principles that enable \pud{}. 
Each DRAM module consists of 8–16 chips, and each chip contains several independent DRAM banks~\circleds{2a} capable of serving requests in parallel.
A bank is composed of multiple mats (subarrays\rncami{~\cite{salp,seshadri-micro-2013, lee.hpca13, seshadri-micro-2017, chang-hpca-2016, chang.hpca14, lee.hpca15, seshadri-ieeecal-2015, chargecache, lee.sigmetrics17}})~\circleds{2b}, each composed of cell arrays~\circleds{2c} organized as rows and columns. 
The mat selector\rncami{~\cite{oliveira2024mimdram}}~\circleds{2d} connects a chosen mat to global bitlines and shared sense amplifiers, while local row decoders~\circleds{2e} activate local wordlines~\circled{2f} to access stored bits. 
The activated row is latched in a row buffer~\circleds{2g}. The global wordlines~\circleds{2h} coordinate access across mats. Each DRAM cell~\circled{2i} stores a single bit using a capacitor-transistor pair. 

\begin{figure}[t]
    \centering    
    \includegraphics[width=0.9\linewidth]{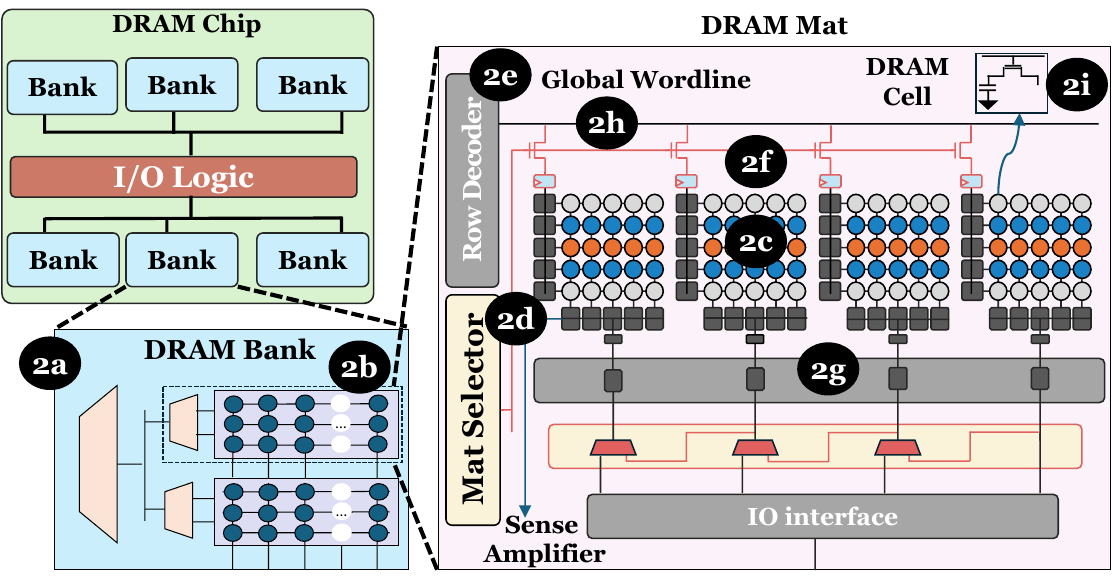}
    \caption{Overview of processing-using-DRAM}
    \label{fig:processing_using_ssd_dram}
\end{figure}

\rncami{Processing using DRAM (PuD) techniques \rncamii{leverage} DRAM's intrinsic operational principles to perform computation directly within memory arrays. 
Prior works \rncamii{(e.g.,~\cite{seshadri-micro-2017, hajinazar2021simdram, oliveira2024mimdram, besta2021sisa, seshadri-arxiv-2019, li-micro-2017, seshadri-micro-2013, seshadri-arxiv-2016-pum, deng-dac-2018, xin-hpca-2020, gao-micro-2019, seshadri-ieeecal-2015, yuksel2024functionally, yuksel2025dram, yuksel2024simultaneous, mutlu2025memory, mutlu2024memory, lenjani_fulcrum_2020, silentpim, park2024attacc, leitersdorf2023aritpim, olgun2022pidram, olgun2023dram, peng_chopper_2023, gao2022fracdram, roy-2021-jetcas})} show that careful orchestration of ACT (row activation) and PRE (precharge) commands enables a range of primitive operations (e.g., AND, OR, NOT, MAJORITY) in a highly parallel manner within the DRAM chips. 
RowClone~\cite{seshadri-micro-2013} demonstrates that consecutively activating two rows within the same DRAM subarray enables fast bulk data copy and initialization by transferring data through shared sense amplifiers. 
Ambit~\cite{seshadri-micro-2017} exploits triple-row activation to compute the bitwise MAJORITY (AND and OR by extension) function, and implements bitwise NOT using sense amplifiers, with DRAM circuit modifications.  

Building on these techniques, SIMDRAM~\cite{hajinazar2021simdram} introduces a software-hardware co-designed framework that maps arbitrary logic circuits composed of AND, OR, and NOT gates onto the Ambit substrate. SIMDRAM enables complex operations (e.g., multiplication, addition, convolution) to be \rncamii{performed} entirely within the DRAM chips. 
MIMDRAM~\cite{oliveira2024mimdram} improves programmability and applicability of the Ambit substrate by (1) supporting finer-granularity operations than full-row execution, and (2) providing compiler support that transparently transforms applications to exploit bulk-bitwise execution.
\rncamiv{Recent studies on commercial-off-the-shelf (COTS) DRAM chips\rncamv{~\cite{mutlu2025memory, mutlu2024memory, yuksel2025dram, yuksel2025columndisturb, yuksel2025pudhammer, yuksel2024simultaneous, yuksel2024functionally, olgun2023dram, gao-micro-2019, gao2022fracdram, olgun2021quactrng, kim-hpca-2019, kim-hpca-2018}} demonstrate that existing DRAM chips are capable of performing all of these operations by violating specific timing parameters in the memory controllers, without modifying the DRAM chip and the DRAM interface.}

Compared to execution on SSD controller cores, \pud{} \rncamii{enables} executing computations in DRAM chips, and significantly improves performance and energy efficiency. However, \pud{} operates on data in the DRAM, which requires transferring operands from flash chips via bandwidth-limited flash channels. These data transfers introduce additional latency and flash channel contention. 
}

\head{In-Flash Processing (IFP)}
\fig~\ref{fig:in_flash_processing} shows the organization of a NAND flash chip and the principles that enable computations inside the flash array. 
A NAND flash chip~\circleds{3a}\rncamii{~\cite{stoica2019understanding, grupp2009characterizing, cai-procieee-2017, yucai-thesis, nadig2023venice, dirik2009performance, micheloni2013inside, cai.bookchapter18.arxiv, cai-insidessd-2018, cai-hpca-2017, cai-date-2012, cai-dsn-2015, cai-iccd-2012, cai-date-2013, cai-hpca-2015, cai-iccd-2013, cai-inteltechj-2013, cai-sigmetrics-2014}} contains 1-4 independent dies~\circleds{3b}, each capable of operating concurrently.
\rncami{Each flash die has multiple planes~\circleds{3c} (typically 1 to 4) that share the flash die's peripheral circuitry for row decoding, sensing, and programming. Each plane holds thousands of blocks~\circleds{3d}.}
Each block consists of hundreds of pages. Each page corresponds to a row of flash cells connected to a single wordline in a block. Read/write operations are performed at page granularity (e.g., 16KB). 
Planes within the same die operate concurrently (\rncami{enabling} multi-plane operations), but only when accessing pages at the same offset. 
Vertically stacked flash cells (typically \rncami{consisting of 24-1024 cells}) connected in series form NAND strings~\circleds{3e} that connect to bitlines (BLs). Multiple NAND strings across different BLs constitute a sub-block~\circled{3f}. 
\rncami{A flash cell}~\circleds{3g} \rncami{stores} bits \rncami{in terms of its} threshold voltage: \rncami{a cell in the programmed state (0) has a high threshold voltage, whereas an erased cell (1) has a low threshold voltage. During reads, a programmed cell behaves as an open switch, and an erased cell operates as a resistor.}

\begin{figure}[b]
\centering
\includegraphics[width=0.95\linewidth]{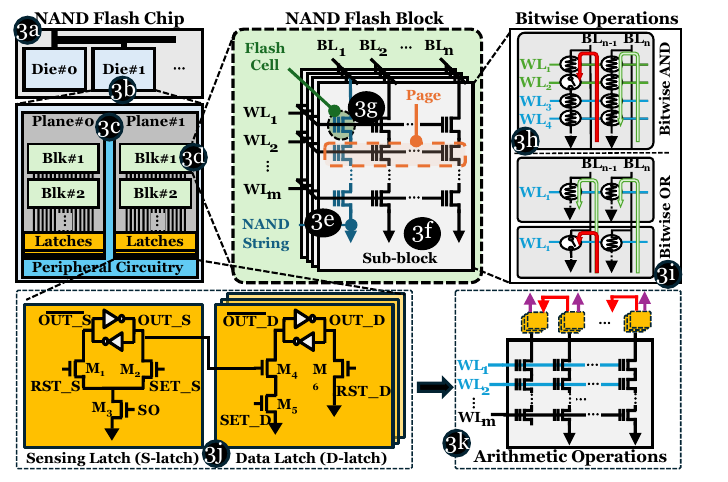}
\caption{Overview of in-flash processing}
\label{fig:in_flash_processing}
\end{figure}

IFP techniques\rncami{~\cite{park2022flash, gao2021parabit, chen2024aresflash, kabra2025ciphermatch, kim2025crossbit, kang-tc-2021, chen2024search, wong2024tcam}} exploit the flash chip's structure to perform bulk bitwise and arithmetic operations within the flash chips. 
Flash-Cosmos~\cite{park2022flash} enables bitwise AND~\circleds{3h} (OR~\circled{3i}) by simultaneously activating multiple wordlines within (across) blocks.
Several works~\cite{chen2024aresflash,kabra2025ciphermatch,gao2021parabit} control the latches (S-latch, D-latch)~\circleds{3j} in the page buffer within the flash die's peripheral circuitry to enable more complex operations such as addition and multiplication~\circleds{3k}.
These techniques exploit both bit-level and array-level parallelism within the SSD to enable efficient in-place computation. 
While \ifp{} eliminates data movement overhead and enables high parallelism, it supports a limited set of operations and is \rncami{therefore} less suitable for workloads that require complex computations.

\section{Motivation}
\rncamiii{We demonstrate} the need for an effective offloading policy for SSD-based NDP. We (1) describe the limitations of NDP techniques proposed for SSDs, (2) present a case study that demonstrates the need for offloading computations across \rncamiii{multiple} SSD computation resources, and (3) quantitatively demonstrate that prior offloading techniques proposed for other NDP architectures fall short when adapted to SSDs.

A modern SSD contains multiple heterogeneous computation resources, including general-purpose embedded controller cores, SSD-internal DRAM, and NAND flash chips (see \S\ref{subsec:background_ndp}). Each computation resource enables an NDP paradigm with different strengths and limitations. We describe the strengths and limitations of each of these computation paradigms in \S\ref{subsec:background_ndp}.

\head{Limitations of SSD-based NDP Techniques}
A large body of prior work \rncamiii{(e.g., \cite{park2016storage,   kabra2025ciphermatch, mailthody-micro-2019, kim-fast-2021, kang-msst-2013, torabzadehkashi-pdp-2019, seshadri-osdi-2014, wang-eurosys-2019,acharya-asplos-1998,keeton-sigmod-1998, wang2016ssd, koo-micro-2017, tiwari-fast-2013, tiwari-hotpower-2012, boboila-msst-2012, bae-cikm-2013,torabzadehkashi-ipdpsw-2018, pei-tos-2019, do-sigmod-2013, kim-infosci-2016, riedel-computer-2001,riedel-vldb-1998, liang-atc-2019,cho-wondp-2013, jun2015bluedbm, lee-ieeecal-2020, ajdari-hpca-2019,liang-fpl-2019,jun-hpec-2016, kang-tc-2021, kim-sigops-2020, lee2022smartsage, li2023ecssd, ruan2019insider, wang2016ssd1, li-atc-2021, jeong-tpds-2019, mao2012cache, gouk2024dockerssd, ghiasi_megis_2024, ghiasi2022genstore, kang-micro-2021, yavits2021giraf, kim2023optimstore, lim-icce-2021, narasimhamurthy2019sage, jun-isca-2018, fakhry2023review, gu-isca-2016, yang2023lambda, jo2016yoursql, chen2025reis, park2022flash, gao2021parabit, chen2024aresflash, soysal2025mars, lee2017extrav, kim2025crossbit, chen2024search, wong2024tcam, pan2024instattention, choi2015energy, okafor2023fusing, pan2024instinfer, istvan2017caribou, choi2020flash, chun2022pif})} proposes SSD-based NDP techniques. These techniques have two key limitations.

First, these techniques operate largely in isolation and map specific portions of an application to \emph{only} one or two SSD computation resources, which prevents them from exploiting the SSD's full computational potential.
For instance, Active Flash~\cite{tiwari-fast-2013} offloads data analytics kernels to SSD controller cores (ISP), and Flash-Cosmos~\cite{park2022flash} accelerates bulk bitwise operations directly within NAND flash chips (IFP).
Our motivational experiments in \S\ref{subsec:motivation_effectiveness} on six real-world workloads show that ISP and IFP, in isolation, underperform by 1.8$\times$ and 1.9$\times$ on average compared to an offloading approach~\cite{ghiasi2022alp} that leverages \rncamiii{\emph{multiple}} SSD computation resources.  

Second, these techniques are typically application-specific and \emph{not} programmer-transparent. 
For example, MARS~\cite{soysal2025mars} accelerates raw signal genome analysis\rncamiii{~\cite{firtina2023rawhash, firtina2024rawhash2, firtina2024rawsamble, lindegger2024rawalign, eris2025rawbench}} by performing computations within the SSD-internal DRAM and the SSD controller. MARS requires the programmer to explicitly identify offloadable sections and manage data placement \rncamiii{and code scheduling} to enable in-situ computations. The need for such programmer intervention limits \rncamiii{generality and ease of adoption in modern storage and system stacks}. 

\subsection{Case Study on Offloading Computations in SSD \label{subsec:motivation_casestudy}}
Offloading computation across multiple SSD computation resources is key to exploiting the SSD's full computational potential. To study how different workloads interact with SSD computation resources, we conduct a case study that examines performance bottlenecks (i.e., data movement or computation) across \rncamiv{three execution models}\footnote{\rncamiv{We exclude \pud{} from this case study because \pud{} incurs \rncamiv{similar} data-movement overheads as ISP, since data must be transferred from flash chips to DRAM over bandwidth-limited flash channels.}} 
(see \S\ref{sec:methodology} for our experimental methodology):
(1) Outside-storage processing (OSP) on host CPU or GPU, 
(2) In-storage processing (\isp{}) on SSD controller cores, and 
(3) In-flash processing (\ifp{}) within flash chips. 
\fig~\ref{fig:motivation_casestudy} presents our analyses for three categories of workloads, I/O-intensive, more compute-intensive, and mixed workloads.

\head{I/O-Intensive Workloads} For workloads dominated by data transfer \rncamiv{(e.g., databases~\cite{Oracle, li-sigmod-2013}, bitmap indices~\cite{fastbit, redis-bitmaps, chan1998bitmap, seshadri-ieeecal-2015})}, OSP is limited by the host-SSD data movement. 
\isp{} outperforms OSP by 30\% by reducing host-SSD data movement, but \isp{}'s performance is limited by the significant data movement between the flash chips and the SSD controller via bandwidth-limited flash channels. 
\rncamiii{IFP outperforms OSP by 70\% by executing computations directly within the flash chips and significantly reducing the data movement}. 
Naively combining IFP and ISP reduces performance by 15\% compared to IFP, because the additional data movement overhead between flash chips and the controller offsets the benefits of IFP’s \rncamiii{highly} parallel computation capability.

\head{More Compute-Intensive Workloads} 
For more compute-intensive workloads (e.g., encryption~\cite{daemen1999aes}, matrix multiplication~\cite{Karpathy}), ISP outperforms OSP by 20\% by avoiding host-SSD transfers, but ISP's limited SIMD parallelism constrains the throughput. 
IFP outperforms both OSP and ISP (50\% improvement over OSP) by exploiting the massive bit- and array-level parallelism of flash arrays. 
Combining IFP and ISP provides additional performance gains \rncamiii{(28\% over IFP)} because: (i) ISP handles control-intensive operations and complex operations that IFP does not support, and (ii) IFP accelerates bulk data-parallel computation. This result demonstrates that leveraging multiple computation resources mitigates the limitations of individual computation paradigms.

\head{Mixed Workloads} 
For workloads that involve both computation and data movement \rncamiv{(e.g., aggregation~\cite{soysal2025mars}, sorting~\cite{vermij2017sorting, stonebraker2013voltdb, soysal2025mars, mirzadeh2015sort, quero2015self, lee2016activesort, salamat2021nascent, myung2020efficient, salamat2022nascent2, lee2015external, wu2015data, wang2016ssd})}, OSP is bottlenecked by host-SSD data transfers.  ISP outperforms OSP by \rncamiii{25\%} by reducing the host-SSD communication, but ISP is constrained by the limited SIMD parallelism in the SSD controller cores.
\rncamiii{IFP outperforms OSP by 53\% by significantly reducing the data movement while exploiting the massive bit- and array-level parallelism of flash arrays.}
Combining IFP and ISP improves performance by \rncamiii{40\% over IFP}, as IFP accelerates bulk data-parallel computations and ISP efficiently executes control-intensive operations and complex operations that IFP does not support.

\begin{figure}[t]
\centering
\includegraphics[width=0.85\linewidth]{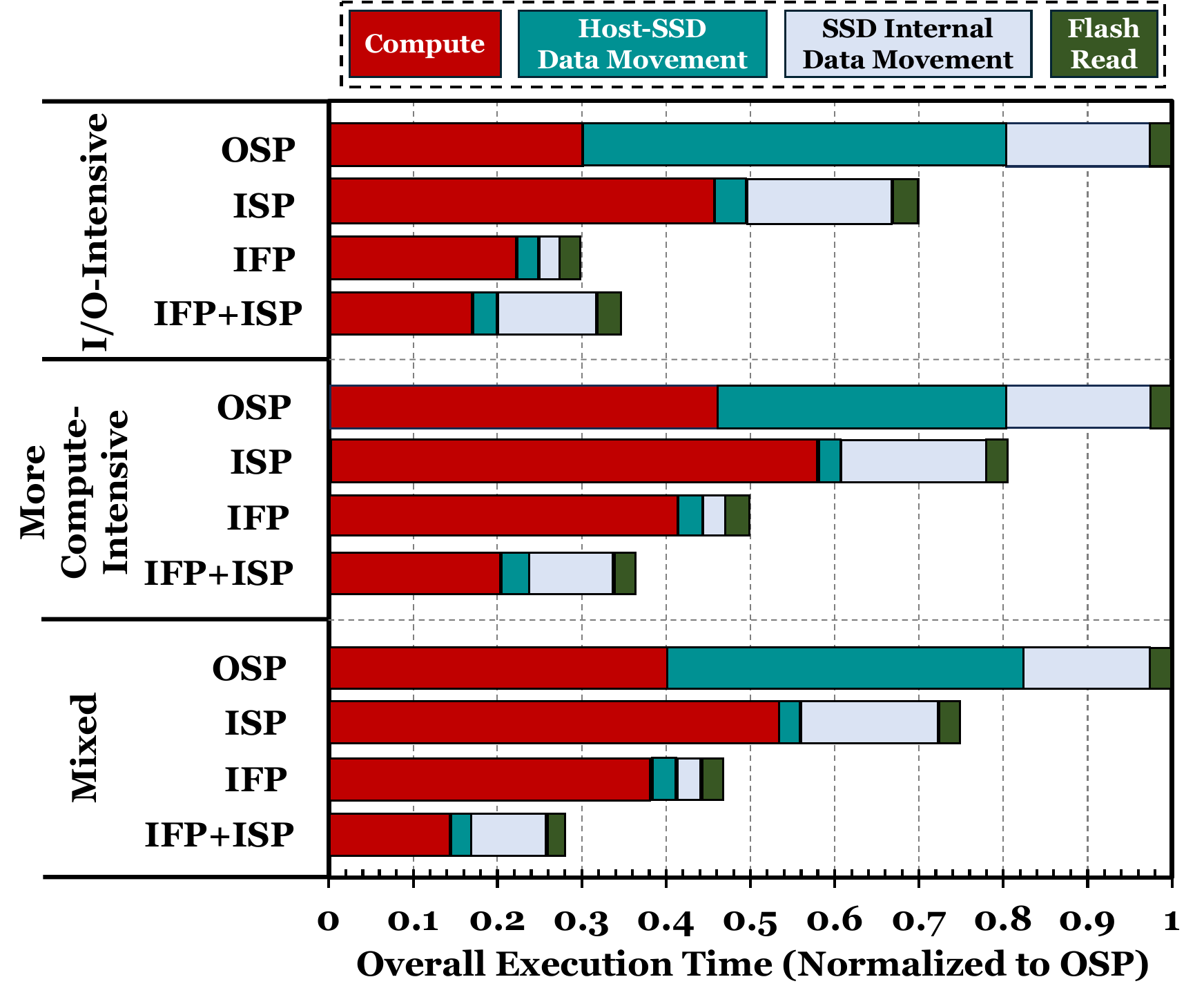}
\vspace{-0.5em}
\caption{Case \rncamiii{study} on offloading computations across computation resources. \rncamiii{Overall execution time of OSP, ISP, IFP and IFP+ISP normalized to \rncamiv{that of} OSP. Lower is better.}}
\label{fig:motivation_casestudy}
\end{figure}

This case study provides three key insights.
First, no single execution model consistently performs well across different workloads. IFP is effective for I/O-intensive workloads where data movement dominates. Combining ISP and IFP is most effective for \rncamiii{the} more compute-intensive and mixed workloads \rncamiii{we examine}. Our case study demonstrates that relying \emph{only} on a single computation resource leads to suboptimal performance across diverse workloads.
Second, performance bottlenecks shift across execution models. OSP's performance is limited by host-SSD data movement. ISP's performance is constrained by its limited SIMD parallelism, and IFP by its limited operation set. These bottlenecks are workload-dependent and can vary \rncamiii{across} different execution phases of the same workload.  
Third, combining multiple SSD computation resources provides benefits, but only when done judiciously.
Naively combining ISP and IFP does not always provide performance gains \rncamiii{and can reduce performance compared to each alone}. In I/O-intensive workloads, combining ISP and IFP leads to additional inter-resource data movement.

This case study demonstrates that effective SSD-based NDP requires fine-grained, workload- and system-aware \rncamiii{scheduling} across multiple heterogeneous SSD computation resources.

\begin{figure*}[t]
\centering
\includegraphics[width=0.95\textwidth]{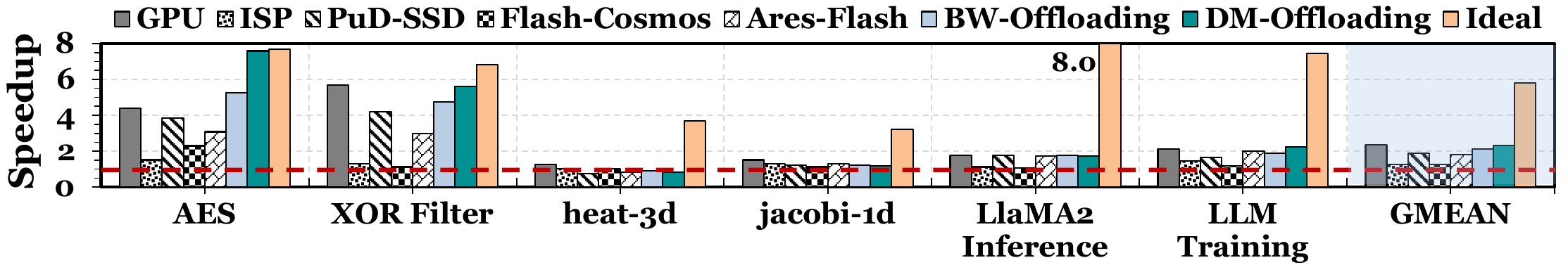}
\caption{Speedup of GPU, \isp{}, \pud{}, \fc{}, \af{}, \bw{}, \dm{} normalized to \cpu{}.}
\label{fig:motivation_prior_offloading}
\end{figure*}

\subsection{Effectiveness of Prior Offloading Approaches \label{subsec:motivation_effectiveness}}
\head{NDP Offloading Techniques} To our knowledge, \emph{no} prior work \rncamiii{proposes} \rncamiii{\emph{general-purpose, application-transparent}} offloading of computations across \rncamiii{\emph{multiple}} heterogeneous \rncamiii{\emph{SSD}} computation resources. 
Several prior techniques (e.g.,~\cite{hsieh.isca16, yang2023lambda, ghiasi2022alp, wolski2008using, hadidi2017cairo, wu2020tuning, alsop2024pim, kim2017toward, wei2022pimprof,jiang20243}) explore partitioning and mapping application code segments for execution between host and NDP units near main memory (e.g., 3D stacked memory with general-purpose cores in its logic layer). These techniques make offloading decisions using a narrow set of system-level characteristics, and do not account for the heterogeneity of SSD computation resources.   
We categorize these techniques into two classes: (1) \textbf{\bw{}}: techniques (e.g.,~\cite{hsieh.isca16, yang2023lambda, wolski2008using,  hadidi2017cairo, wu2020tuning, alsop2024pim}) that make offloading decisions based on bandwidth utilization of the host and NDP units, and (2) \textbf{\dm{}}: techniques (e.g.,~\cite{ghiasi2022alp, kim2017toward, wei2022pimprof,jiang20243}) that prioritize minimizing data movement cost when executing application code across host and NDP units. 

\head{Methodology}
To evaluate the effectiveness of these prior offloading models when adapted to SSD-based NDP, we extend both \bw{} and \dm{} models to utilize all three SSD computation resources.
We implement these approaches in our in-house event-driven simulator and evaluate them across six real-world workloads (see \S\ref{sec:methodology} for our evaluation methodology and workload characteristics).
We compare their performance against:
(1) host CPU and GPU, 
(2) standalone NDP techniques (see \S\ref{subsec:background_ndp}) - \isp{}, \pud{}, \fc{}\rncamiii{~\cite{park2022flash}}, and \af{}\rncamiii{~\cite{chen2024aresflash}}, and 
(3) an \ideal{} approach that assumes (i) no resource contention (for both computation resources and shared channels), (ii) zero data movement latency to move the operands to the target computation resource, and (iii) the selection of a computation resource that provides the lowest computation latency \rncamiii{for each instruction}.

\head{Performance Analysis} \fig~\ref{fig:motivation_prior_offloading} shows the speedup of GPU, \isp{}, \pud{}, \fc{}, \af{}, \bw{}, \dm{} and \ideal{} normalized to \cpu{}.

We make four key observations.
First, \dm{} is the best-performing prior NDP offloading technique with an average speedup of 2.3$\times$ over CPU, \rncamiii{0.98$\times$} over GPU, 1.8$\times$ over \isp{}, 1.3$\times$ over \pud{}, 1.9$\times$ over \fc{}, 1.3$\times$ over \af{}, and 1.1$\times$ over \bw{}. 
\dm{}'s benefits come from offloading computation to multiple SSD computation resources based on data movement cost reduction. 
Second, \dm{} falls short of an \ideal{} policy's performance by 2.5$\times$ on average \rncamiii{(and by up to 4.6$\times$ for LLM inference and training)}. This is because \dm{} relies on a \rncamiii{limited} offloading model that frequently favors \ifp{} to reduce data movement costs. However, repeated offloading to flash chips (i.e., IFP) introduces queueing delays and resource contention, which reduces IFP's effectiveness. 
Third, \bw{} achieves an average speedup of 2.1$\times$ over CPU but underperforms \dm{} by 11\%. \bw{} prefers \rncamiii{lightly-loaded} computation resources based on bandwidth utilization without considering the cost of moving operands across the computation resources. This \rncamiii{result} shows that bandwidth awareness alone is insufficient for SSD-based NDP, where internal data movement and contention can \rncamiii{greatly affect} performance.
Fourth, GPU has comparable \rncamiii{average} performance to \dm{}, and even outperforms \dm{} in highly data-parallel workloads (e.g., heat-3d and jacobi-1d). These workloads benefit from GPU's massive SIMD parallelism and wide vector processing units. 

We conclude that prior offloading techniques consider limited factors, which leads to \rncamiii{suboptimal} performance compared to the \ideal{} offloading policy. An effective offloading policy must jointly consider computation capability, data locality, and resource contention across heterogeneous SSD computation resources.

\textbf{Our goal} is to enable programmer-transparent near-data processing in SSDs that (1) schedules and coordinates computation across multiple heterogeneous SSD computation resources, (2) makes offloading decisions that are aware of both workload characteristics and dynamic system conditions, and (3) improves the performance and energy efficiency of a wide range of applications.

\section{\namePaper{} \label{sec:mechanism}}
\subsection{Overview}
We propose \namePaper{}, a \emph{general-purpose programmer-transparent} NDP framework that dynamically offloads fine-grained computations (at instruction granularity) to \emph{multiple} heterogeneous SSD computation resources. 
\namePaper{} consists of two key steps: 
(1) \textbf{Compile-time preprocessing}, where \namePaper{} identifies offloadable code regions (e.g., loops with computations) and transforms them into SIMD operations that match the SSD's internal bit-level parallelism (see \S\ref{subsec:mechanism_offline_preprocessing}), and 
(2) \textbf{Runtime dynamic offloading}, where \namePaper{} (i) selects the most suitable SSD computation resource to execute each vector operation using a holistic cost function, which \rncamiv{takes into account} six key factors: operation type, operand location, data dependencies, resource utilization, data movement costs, and expected computation latencies, (ii) translates each vector operation to the native ISA of the chosen resource, and (iii) dispatches the instruction to the chosen resource's execution queue (see \S\ref{subsec:mechanism_runtimeoffloading}).

\subsection{Design Challenges for an SSD Offloading Policy}
Offloading across multiple SSD computation resources is non-trivial and presents three key challenges. 
First, SSD computation resources exhibit heterogeneity in computation capabilities, parallelism, and data-access latencies. This architectural heterogeneity makes it challenging to determine a suitable computation resource for each instruction. 
Second, the shared buses (e.g., DRAM bus, flash channels) in an SSD are prone to contention (e.g.,~\cite{nadig2023venice, kim2022networked}). Offloading decisions that cause frequent data transfers over these shared buses can reduce the benefits of NDP, making \rncamiv{SSD-}internal data movement a critical bottleneck.
Third, an effective offloading policy must (1) provide performance benefits across a wide range of workloads, (2) balance utilization of all the available computation resources, and (3) reduce the programmer's burden of manually identifying offloadable code sections and mapping them to suitable computation resources.

\subsection{Detailed Design of \namePaper{}}
\fig~\ref{fig:design_overview} shows \namePaper{}'s high-level design. We describe \namePaper{}'s components and its execution flow in detail. 

\begin{figure}[b]
    \centering
    \includegraphics[width=\linewidth]{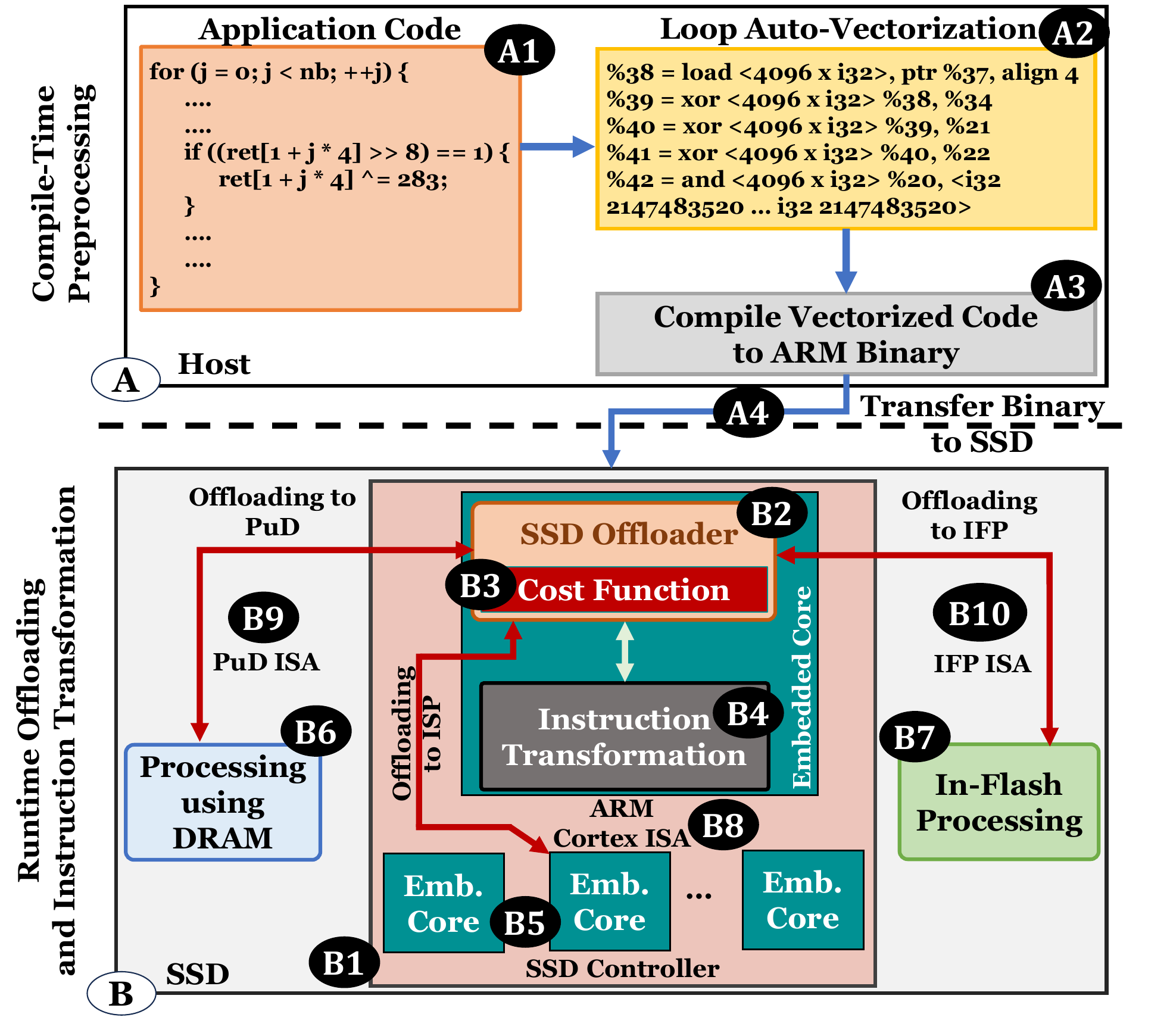}
    \caption{Design Overview of \namePaper{}.}
    \label{fig:design_overview}
\end{figure}
    
\subsubsection{Compile-Time Preprocessing in the Host\label{subsec:mechanism_offline_preprocessing}}
\namePaper{} performs a \emph{one-time} offline preprocessing~\whitecircle{A} of the application for NDP in an SSD. 
This phase consists of \rncamiv{three} steps: \rncamiv{(1) loop auto-vectorization, (2) vectorized code compilation, and (3) binary transfer}. 
\rncamiv{Compile-time} preprocessing (1) keeps the runtime offloading decisions lightweight and \rncamiv{thus,} reduces the burden on the SSD controller, and (2) enables near-data processing with \rncamiv{small} changes to the storage and system stacks.

\head{Loop Auto-Vectorization}
\namePaper{} runs a custom compiler (e.g., LLVM \cite{llvm}) pass on the application code~\circled{A1} to identify offloadable code regions (e.g., loops with computations). 
We compile applications with \rncamiv{the} Clang compiler~\cite{clang}, using custom flags to enable loop auto-vectorization. 
Loop auto-vectorization is a well-known compiler optimization technique that preserves programmer transparency, and is used in many prior DRAM-based NDP techniques \rncamiv{(e.g., ~\cite{hajinazar2021simdram,oliveira2024mimdram,ahmed2019compiler, de2025proteus, devic2022pim,singh2025scalable, khadem2023vector})}.  
We perform loop auto-vectorization~\circled{A2} using LLVM compiler toolchain (version 12.0.0)~\cite{llvm} to transform scalar instructions into wide vector (i.e., SIMD) operations that match \rncamiv{the SSD-internal} parallelism. 

\namePaper{} makes three customizations to the loop auto-vectorization process to accommodate SSD-specific constraints and heterogeneous computation resources. 
First, \namePaper{} compiles the application code with flags ({\small\texttt{-O3 -g -mllvm -force-vector-width=4096 -force-vector-interleave=1 -Rpass-analysis=loop-vectorize -Rpass=loop-vectorize}}).
The {\small\texttt{-force-vector-width=4096}} flag configures the vector width to 4096 for 32-bit operands (16KiB in total), which aligns each vector to a typical NAND flash page.
This (i) simplifies data movement across computation resources, (ii) avoids misaligned accesses, and (iii) matches FTL's logical-to-physical (L2P) mapping granularity. \rncamiv{While this vector width is optimized for \ifp{}, other SSD computation resources (i.e., \pud{} and \isp{}) support smaller vector widths. \namePaper{}'s runtime SSD offloader handles this mismatch at runtime (see \S\ref{subsec:mechanism_runtimeoffloading})}.
The {\small\texttt{-force-vector-interleave=1}} flag controls loop unrolling (i.e., how many iterations are processed simultaneously) to enable instruction-granularity offloading.

Second, not all loops are fully vectorizable due to control flow or data dependencies (see \S\ref{sec:discussion}). \namePaper{} leverages \rncamiv{\emph{partial}} vectorization (e.g., strip-mining~\cite{baghsorkhi2016flexvec}) so that partially vectorizable code regions can still benefit from SIMD execution in SSDs. This is essential for workloads with mixed data dependencies (e.g., LLM attention kernels~\cite{Karpathy, pan2024instinfer, pan2024instattention, lee2025aif}). 

Third, the custom compiler pass records lightweight metadata (e.g., instruction type, operand pointers, element sizes, vector length) for each vector operation. 
During the preprocessing step, \namePaper{} takes the application's LLVM intermediate representation (IR) as input and generates an optimized IR that contains SIMD instructions and metadata. 
Embedding the metadata in the optimized IR at compile time reduces runtime scheduling latency and preserves programmer transparency.

\head{Vectorized Code Compilation and Transferring the Binary\label{Workload Compilation}}
\namePaper{}'s compile-time preprocessing step converts the optimized IR to a binary executable file based on ARM ISA~\circled{A3}, and transfers the binary~\circled{A4} to the SSD. 
We select \rncamiv{the} ARM ISA because SSD controllers \rncamiv{(e.g.,\cite{marvellsc5bravera, armssdcontroller, cai-procieee-2017, micheloni2013inside, cheong2018flash, hatanaka2012nand, liao2012multi, park2006high, do2019improving, dirik2009performance, takeuchi2009novel,kim2021decoupled, kim2022networked})} typically \rncamiv{use} ARM cores (e.g.,~\cite{arm-cortexR5, arm-cortexR8, armssdcontroller}).
To transfer the ARM binary from the host to the SSD, \namePaper{} repurposes existing NVMe admin commands~\cite{nvme} for firmware update, {\small \texttt{fw-download}} and {\small \texttt{fw-commit}} (See \S\ref{subsec:system_modifications} for details). 

\subsubsection{Runtime Offloading and Instruction Transformation\label{subsec:mechanism_runtimeoffloading}}
At runtime~\whitecircle{B}, \namePaper{} dynamically identifies the most suitable SSD computation resource for each SIMD instruction in the binary, transforms the instruction to the target resource's ISA, and dispatches the transformed instruction to the target resource's execution queue. 
\namePaper{} performs this step entirely inside the SSD controller~\circled{B1} for two key reasons. First, the SSD controller has real-time knowledge of the current state of all SSD computation resources (e.g., execution queue occupancy, flash channel utilization, data location), which enables timely offloading decisions based on current system conditions.
Second, making offloading decisions within the SSD avoids changes to the host-side storage stack and preserves programmer transparency.

\rncamiv{\namePaper{}'s runtime offloading is tightly coupled with its data mapping and internal data transfer policies (see \S\ref{subsec:system_modifications}). This allows offloading decisions to account for operand placement and internal data movement during instruction execution.}
The runtime offloading phase involves three components: (1) the SSD offloader~\circled{B2}, (2) a holistic cost function~\circled{B3}, and (3) instruction transformation~\circled{B4}. 

\head{SSD Offloader} \namePaper{}'s SSD offloader~\circled{B2} is responsible for mapping each vectorized instruction in the binary executable to the most suitable SSD computation resource at runtime and executing them.
This unit operates alongside the flash translation layer (FTL) within the SSD controller.
At runtime, \namePaper{} executes the transferred ARM executable~\circled{A4} on one of the several \rncamiv{embedded} cores~\circled{B5} present in the SSD controller. 
The SSD offloader dynamically selects the most suitable computation resource by evaluating a holistic cost function~\circled{B3} (see Cost Function in \S\ref{subsec:mechanism_runtimeoffloading}).  
Based on the cost function's decision, the SSD offloader selects one of three NDP paradigms (see \S\ref{subsec:background_ndp} for detailed background): 
(1) processing using the controller cores (\isp{})\footnote{\rncamiv{We leverage one of the several SSD controller cores to execute offloaded computations. We use the other controller cores for latency-critical tasks, including FTL functions, host communication, and \namePaper{}'s offloading and instruction transformation tasks.}}~\circled{B5}, 
(2) processing-using-DRAM in SSD (\pud{}) ~\circleds{B6}, and 
(3) in-flash processing (\ifp{})~\circleds{B7}.

\rncamiv{At compile time, \namePaper{} vectorizes instructions to match the NAND flash page size (i.e., 16KiB). However, this vector width may not always align with other SSD computation resources, such as SSD DRAM (e.g., 8KiB) or the controller cores (e.g., 32B), which support smaller SIMD widths. 
To address this mismatch, the SSD offloader dynamically adjusts the vector length by splitting the original operation into multiple smaller operations that suit the target computation resource's architectural constraints. 
For example, a 4096-element vector operation is decomposed into several \rncamv{2048-element} \rncamv{(or smaller)} sub-operations for execution in DRAM or controller cores.}

When the offloader selects an SSD computation resource to execute an instruction, \namePaper{} performs instruction transformation~\circleds{B4} (see Instruction Transformation Unit in \S\ref{subsec:mechanism_runtimeoffloading}) to translate the SIMD instruction to the native ISA of the target computation resource. 
After instruction transformation, the SSD offloader dispatches the instruction to the target computation resource's execution queues (see \S\ref{sec:background}).
\namePaper{} enables concurrent utilization of SSD's multiple computation resources by leveraging their execution queues. For example, while flash chips are performing computations, the SSD DRAM or SSD controller core can execute other independent instructions simultaneously. This overlap allows \namePaper{} to exploit parallelism across resources and mitigate performance bottlenecks during execution.

\namePaper{}'s offloading unit incurs minimal runtime overhead because the cost function evaluation and instruction transformation are lightweight operations (see \S\ref{subsec:overhead}). 
The more compute-intensive offline preprocessing phase (see \S\ref{subsec:mechanism_offline_preprocessing}) is performed on the host CPU during compile time. 

\head{Cost Function} 
\namePaper{} uses a holistic cost function to select the SSD computation resource with the lowest offloading cost for each instruction. The cost function uses six key features (see Table~\ref{tab:offloader_parameters}) from the application and current system state. 

\noindent (1) \textit{Operation Type}.
This feature indicates the type of computation (e.g., bulk-bitwise, arithmetic, predication) in the vectorized instruction.
\rncamiv{SSD computation resources differ significantly in the operations they currently support. \isp{} (e.g., \cite{arm-cortexR8, m-profile}) supports $\sim$300 ISA instructions. \pud{} (e.g., \cite{oliveira2024mimdram, hajinazar2021simdram, de2025proteus}) supports 16 operations, including arithmetic, predication, and relational operations. \ifp{} (e.g., \cite{kabra2025ciphermatch, park2022flash, chen2024aresflash}) supports \rncamv{nine operations: six bitwise operations, and three arithmetic operations.}}
\namePaper{} embeds the operation type in the optimized IR at compile time. 
At runtime, \namePaper{} estimates the execution latency of an instruction in each computation resource based on the operation type.

\begin{table}[b]
\centering
\setlength{\tabcolsep}{2pt}
  \resizebox{\linewidth}{!}{%
\begin{tabular}{|c|l|}
\hline
\textbf{Feature} &
\makecell[c]{\textbf{Description}} \\
\hline
\hline
\textbf{Operation Type} & 
\begin{tabular}{l}
\rncamiv{Type of operation} \\ \rncamiv{(e.g., bulk-bitwise, arithmetic, predication)}
\end{tabular}
\\
\hline
\textbf{Operand Location} & 
\begin{tabular}{l}
\rncamiv{Current location of the operand} \\ \rncamiv{(flash chips or SSD DRAM)}
\end{tabular}
\\
\hline
\textbf{Data \rncamiv{Dependence} Delay ($delay_{dd}$)} &
\begin{tabular}{l}
Delay for operand availability 
\end{tabular}
\\
\hline
\begin{tabular}{c}
\textbf{\rncamiv{Resource Queueing Delay}} \\ ($delay_{queue}$)
\end{tabular}
&
\begin{tabular}{l}
\rncamiv{Delay caused by pending instructions} \\ \rncamiv{in a computation resource's execution queue}
\end{tabular}
\\
\hline
\begin{tabular}{c}
\textbf{Data Movement Latency} \\ ($latency_{dm}$)
\end{tabular}
&
\begin{tabular}{l}
Data movement latencies if the operands \\ are not in the target computation unit
\end{tabular}
\\
\hline
\begin{tabular}{c}
\textbf{Expected Computation Latency} \\ ($latency_{comp}$)
\end{tabular}
&
\begin{tabular}{l}
Estimated instruction execution latency \\ on a computation resource
\end{tabular}
\\ 
\hline
\end{tabular}
}
\caption{Features Used by the Cost Function to Select the SSD Computation Resource.}
\label{tab:offloader_parameters}
\end{table}

\noindent (2) \textit{Operand Location}. 
\namePaper{} tracks each operand's current location \rncamiv{(i.e., flash chips, SSD DRAM)} via \rncamiv{an} L2P mapping table lookup (see \S\ref{sec:background}). At runtime, \namePaper{} uses \rncamiv{the operand location} to estimate data movement latencies when the target computation resource differs from the current operand location. 

\noindent (3) \rncamiv{\textit{Data Dependence} Delay ($delay_{dd}$)}.
This feature captures the time until an instruction's operands become available.
For each instruction, \namePaper{} checks if operands are ready or if the execution must wait for pending computations to complete.
\namePaper{} estimates \rncamiv{dependence-induced} stall time by summing the predicted computation costs of pending instructions that produce the required operands. 
This ensures correct ordering of instructions and avoids pipeline \rncamiv{bubbles}.

\noindent (4) \rncamiv{\textit{Resource Queueing Delay} ($delay_{queue}$)}. 
\rncamiv{\namePaper{} tracks the delay caused by pending instructions in each computation resource's execution queue. This feature reflects the computation resource's current utilization. By incorporating resource queueing delays into the cost function, \namePaper{} aims to reduce resource congestion, which results in better load balancing across the SSD computation resources.}

\rncamiv{\noindent (5) \textit{Data Movement Latencies} ($latency_{dm}$).} 
\namePaper{} estimates the latency cost of moving operands between computation resources (e.g., flash chips to SSD DRAM, SSD DRAM to SSD controller). We precompute the expected data-movement latencies for different operand locations (e.g., SSD DRAM, flash chips), operand sizes, and target computational resources, and store these latencies in the SSD DRAM.
\rncamiv{These latencies capture the data transfer cost over SSD internal interconnects (e.g., flash channels and SSD DRAM bus) under no contention. \namePaper{} does not explicitly model real-time contention for this feature, because accurately tracking utilization across multiple flash channels and the DRAM bus incurs significant runtime overhead.} 
\rncamv{$latency_{dm}$ serves as a static estimate that enables the offloader to avoid decisions that inherently incur internal data movement. \namePaper{} incorporates contention effects on the execution latency when the instruction is executed in the computation resource chosen by the SSD offloader.}

\rncamiv{\noindent (6) \textit{Expected Computation Latency} ($delay_{comp}$).}
Based on profiling data and analytical models (see \S\ref{subsec:methodology_modeling}), \namePaper{} estimates the execution latency of each instruction on every computation resource. This feature captures the differences in computation capabilities and parallelism across computation resources. 

\rncamiv{We calculate the latency of offloading an instruction to each computation resource, $total\_latency\_resource_{i}$, using Eqn.~\ref{eq:overall_latency}:}

\vspace{-1em}
\begin{equation} \label{eq:overall_latency}
\begin{split}
total\_latency\_resource_{i} = latency_{comp}+\\latency_{dm}+ \max(delay_{dd}, delay_{queue})
\end{split} 
\end{equation}
\vspace{-1em}

where $i$ denotes the computation resource (i.e., \isp{}, \pud{}, or \ifp{}), $latency_{comp}$ is the expected computation latency, $latency_{dm}$ is the data movement latency, $delay_{dd}$ and $delay_{queue}$ are the data \rncamiv{dependence delay and resource queueing delay}, respectively. \rncamiv{We calculate the maximum of data-dependence and resource queueing delays because these delays overlap, i.e., an instruction can be executed only when the operands and the computation resource are ready}.

\namePaper{} selects the target computation resource using the cost function in Eqn.~\ref{eq:cost_function}:

\vspace{-1em}
\begin{equation} \label{eq:cost_function}
\begin{split}
offloading\_target = argmin(total\_latency_{ISP}, \\ total\_latency_{PuD\_SSD}, total\_latency_{IFP})
\end{split}
\end{equation}
\vspace{-1em}

For each vectorized instruction, \namePaper{} calculates the latency of executing computations on each resource, and selects the computation resource with the \rncamiv{lowest execution latency}.

\head{Instruction Transformation Unit}
\namePaper{}'s instruction transformation unit transforms each vectorized instruction into the native ISA of the target computation resource. 
\namePaper{} performs this transformation in the SSD controller on the critical path of instruction execution.

For \rncamiv{\isp{}}, we utilize ARM's M-Profile Vector Extension (MVE)~\cite{m-profile} (\circleds{B8} in \fig~\ref{fig:design_overview}) to exploit SIMD execution on ARM cores.
We use the ISA proposed by prior NDP techniques to enable computation on SSD DRAM and flash chips. 
For \rncamiv{\pud{}}, we use ISA extensions (e.g., {\texttt{bbop\_op}} for 2-input arithmetic) from \rncamiv{SIMDRAM~\cite{hajinazar2021simdram}, MIMDRAM~\cite{oliveira2024mimdram}, and Proteus~\cite{de2025proteus}} (\circleds{B9} in \fig~\ref{fig:design_overview}). 
For \rncamiv{\ifp{}}, we leverage the multi-wordline sensing {\texttt{MWS}} primitives from Flash-Cosmos~\cite{park2022flash}, and {\texttt{shift\_and\_add}} from \af{}~\cite{chen2024aresflash} (\circleds{B10} in \fig~\ref{fig:design_overview}). 

\subsection{System Integration\label{subsec:system_modifications}}

\head{Transferring the Binary} We transfer the \rncamiv{ARM binary} from the host to the SSD using existing \rncamiv{NVMe firmware update admin commands~\cite{nvme}}\rncamv{:} {\small\texttt{fw-download}} and {\small\texttt{fw-commit}}. 
To distinguish \rncamiv{between \namePaper{}'s} binary and FTL firmware binary from SSD manufacturers, we extend the firmware update commands with a new flag. The SSD controller interprets this flag to identify the binary and executes it on the controller core.

\rncamiv{\head{Data Mapping and SSD-Internal Data Movement}
We assume that all application data resides in the SSD at the start of application execution, and the application can be executed entirely using the SSD computation resources.
\namePaper{} relies on the FTL to manage data mapping and data transfer across SSD computation resources. 
All data is addressed at logical-page granularity, and the L2P mapping table tracks the current physical location of each page. 
When the SSD offloader selects a target computation resource, \namePaper{} consults the L2P table to locate the operands, and moves the operands \rncamv{to the target \rncamvi{resource if} they are not already resident there.}
We extend the FTL to enforce the data layout constraints of different NDP paradigms. For example, \fc{}~\cite{park2022flash} requires all the operands of a bitwise AND operation to be placed in the pages of the same flash block. Similarly, we adhere to the access and alignment constraints~\cite{mutlu2025memory, mutlu2024memory, yuksel2025dram, yuksel2025columndisturb, yuksel2025pudhammer, yuksel2024simultaneous, yuksel2024functionally, olgun2023dram, gao-micro-2019, gao2022fracdram, olgun2021quactrng, kim-hpca-2019, kim-hpca-2018, de2025proteus} of \pud{} when placing data in DRAM for computation.
We transfer data between flash chips, SSD DRAM, and controller cores using existing read and DMA operations.
}

\head{Host-SSD Communication} 
To enable both host I/O processing and NDP on an SSD, we enable two operating modes: (1) \emph{regular I/O mode}, where host I/O and FTL operations are executed, and (2) \emph{computation mode}, where all computation resources (controller cores, SSD DRAM, and flash chips) are utilized for NDP. 
In computation mode, host I/O traffic is suspended, and  \namePaper{} allocates all the SSD resources for NDP. 
The SSD controller performs maintenance tasks such as garbage collection and wear-leveling in both modes. 
The SSD reverts to regular I/O mode when the host explicitly notifies the SSD controller.

\head{Coherence between SSD Computation Resources}
\rncamiv{To maintain data \rncamiv{coherence} across resources, \namePaper{} employs a lazy coherence mechanism.
A strict coherence policy that synchronizes data after every modification is unsuitable for an SSD as it incurs high latency, increases energy consumption, and accelerates flash wear. 
\namePaper{} synchronizes data across SSD computation resources \emph{only} when: (i) another computation resource requests data, (ii) the computation result must be transferred to the host, and (iii) data must be evicted from SSD DRAM, page buffers or SSD controller registers to reuse these temporary locations for computations, (iv) the FTL initiates garbage collection, and (v) the SSD undergoes a power cycle.
The data remains local to the SSD computation resource until the coherence mechanism performs synchronization.
}

\rncamiv{\namePaper{} maintains coherence at logical-page granularity using lightweight metadata stored in the L2P table in SSD DRAM. For each logical page, the L2P mapping table stores three fields: 
(1) \emph{owner}, the SSD computation resource (i.e., flash or DRAM) that holds the latest version of the page, 
(2) \emph{state}, the page's modification status (i.e., clean, dirty), and 
(3) \emph{version}, \rncamv{a one-byte} monotonically increasing counter to order updates and detect stale copies.\footnote{\rncamvi{A 3-bit counter for \emph{version} ensures correctness in our evaluated workloads because \namePaper{}: (i) enforces a single owner per logical page at any time, and (ii) flushes a page before the counter wraps around. We extend the counter to one byte to support future extensions.}} 
When a computation resource modifies a page, \namePaper{} updates the owner field, sets the state to dirty, and increments the version counter. 
When a dirty page is updated by the same computation resource, \namePaper{} only increments the version number. 
If another computation resource or the host requests the page, \namePaper{} commits the updated page to the NAND flash chips, sets the owner field to flash, marks the state as clean, and resets the version.
}

\head{Failure Handling}
\namePaper{} leverages existing SSD firmware mechanisms for error mitigation. For transient faults (e.g., ECC correction failure~\cite{cai-date-2012, cai-dsn-2015, cai-iccd-2012, cai-hpca-2015, cai-hpca-2017, cai-insidessd-2018}, instruction abort), the SSD controller detects timeouts and invalidates the instruction. \namePaper{} maintains completion and status flags, and the scheduler replays the instruction on another resource using the latest data version. For permanent faults (e.g., \rncamiv{bad blocks~\cite{cai-date-2012, cai-hpca-2017, cai-insidessd-2018, cai-procieee-2017, cai-iccd-2013, cai.bookchapter18.arxiv}}), \namePaper{} relies on existing FTL recovery mechanisms such as bad-block remapping, garbage collection, and wear-leveling\rncamiv{~\cite{cai-date-2012, cai-hpca-2017, cai-insidessd-2018, cai-procieee-2017, cai-iccd-2013, cai-iccd-2012, cai.bookchapter18.arxiv}}.

\subsection{Overhead Analysis \label{subsec:overhead}}
\head{Storage Overhead} 
\namePaper{} maintains a small metadata table in the SSD DRAM to guide the cost function in identifying the target computation resource. 
Table~\ref{tab:offloader_parameters} lists the parameters stored in the metadata table.
We use \rncamiv{two bytes} to store the computation type (e.g., 0x01 represents bitwise AND). 
\rncamiv{We encode operand location using \rncamiv{four} bits (e.g., 0 = flash, 1 = DRAM) to provide support for more operand locations in the future.}
We use two bytes to store the data \rncamiv{dependence} delays (measured in cycles).
\rncamiv{We compute each computation resource's queueing delay\footnote{\rncamv{\namePaper{} tracks the queueing delay at each computation resource by maintaining a running counter of the total execution latency (in cycles) of instructions currently enqueued in the execution queue. When a new instruction is dispatched, \namePaper{} increments the counter by the instruction's estimated execution latency. When the instruction is executed, \namePaper{} decrements the execution latency from the counter.}} as the latency (in cycles) of execution of pending instructions in its execution queue. This requires \rncamiv{four} bytes for each computation resource.}
The data movement and computation latencies (in cycles) each consume \rncamiv{four} bytes. 

\rncamiv{To support instruction transformation, \namePaper{} stores a translation table in the SSD DRAM that maps more than 300 operation types (listed in Table~\ref{tab:offloader_parameters}) to native instructions for each computation resource (i.e., SSD controller cores, SSD DRAM chips, and flash chips). 
Each entry maps the vectorized instruction to its corresponding native instruction, which consumes four bytes. This table incurs a total storage overhead of 1.5 KiB in SSD DRAM.}

\head{Runtime Latency Overhead}
\namePaper{}’s runtime latency overhead includes (1) feature collection latency (i.e., gathering six features \rncamiv{listed} in Table~\ref{tab:offloader_parameters}), and (2) instruction transformation latency (i.e., latency of mapping vector operations to the ISA of the target computation resource).

Runtime feature collection (see Table~\ref{tab:offloader_parameters}) involves five steps. 
First, extracting operand location involves an L2P table lookup, which incurs an average latency of 100 $n$s \rncamv{(30 $\mu$s)} per operand if the L2P mapping entry is in the SSD DRAM (flash chips). 
Second, to measure data \rncamiv{dependence} delays, \namePaper{} adds the predicted computation costs of pending instructions that produce the required operands. This involves tracking pending instructions in each computation resource's execution queue, which incurs \rncamv{1 $\mu$s} per queue on average.
\rncamiv{Third, \namePaper{} tracks resource queuing delays by computing the cumulative execution latency of pending instructions in each computation resource's execution queue. Tracking the queueing delay for each resource incurs \rncamv{1 $\mu$s} on average.}
Fourth, we precompute the expected data-movement latencies for different operand locations, operand sizes, and target computational resources, and store these latencies in the SSD DRAM. A lookup of the expected data-movement latencies incurs 100 $n$s on average. 
Fifth, we precompute expected latencies of computations in each resource and store them in SSD DRAM. Each lookup of expected computation latencies incurs 150 $n$s on average.

\rncamiv{During instruction transformation, \namePaper{} performs a lightweight lookup of a translation table that is stored in the SSD DRAM. 
This translation table lookup incurs an average latency of 300 $n$s}.
Combined with feature collection, the total runtime latency overhead is 3.77 $u$s on average (up to 33 $u$s).
 
\head{Area Overhead} 
\namePaper{} is implemented entirely in the firmware running on the SSD controller, and requires no hardware modifications. 
It builds on the hardware capabilities proposed by prior NDP techniques. 
For processing-using-DRAM, \namePaper{} leverages MIMDRAM~\cite{oliveira2024mimdram}, which incurs an area overhead of 1.11\% on the DRAM array\rncamiv{~\cite{oliveira2024mimdram, mutlu2025memory, mutlu2024memory, yuksel2025dram, yuksel2025columndisturb, yuksel2025pudhammer, yuksel2024simultaneous, yuksel2024functionally, olgun2023dram, gao-micro-2019, gao2022fracdram, olgun2021quactrng, kim-hpca-2019, kim-hpca-2018}}. 
For in-flash processing, \namePaper{} relies on Flash-Cosmos~\cite{park2022flash} and Ares-Flash~\cite{chen2024aresflash}.
\fc{} introduces minimal modifications to the peripheral circuitry of flash chips\rncamiv{~\cite{park2022flash, agrawal2008design, cai-date-2012, cai-iccd-2012, cai-inteltechj-2013, cai-date-2013, cai-iccd-2013, cai-sigmetrics-2014, cai-hpca-2015, cai-dsn-2015, cai-hpca-2017, luo-sigmetrics-2018, cai-procieee-2017, cai-insidessd-2018}}.
\rncamiv{\af{} adds additional transistors and transmission wires to each latch in the flash chip's page buffer, which incurs a 1.5\% area overhead to the flash chip's peripheral circuitry~\cite{chen2024aresflash, agrawal2008design, cai-date-2012, cai-iccd-2012, cai-inteltechj-2013, cai-date-2013, cai-iccd-2013, cai-sigmetrics-2014, cai-hpca-2015, cai-dsn-2015, cai-hpca-2017, luo-sigmetrics-2018, cai-procieee-2017, cai-insidessd-2018}}.

\section{Evaluation Methodology \label{sec:methodology}}
\subsection{Simulator Overview \label{subsec:methodology_simulation}}
To evaluate \namePaper{} and the baselines, we develop an in-house event-driven SSD simulator that (1) builds on the state-of-the-art SSD simulator, MQSim~\cite{tavakkol2018mqsim, mqsim-github}, which is validated against real SSDs, and (2) adds computation models for SSD controller cores, SSD DRAM, and NAND flash chips to support NDP. Table~\ref{tab:eval_config} describes the characteristics of our simulated SSD.

\head{SSD Modeling} 
\namePaper{} leverages MQSim's cycle-\rncamv{level} modeling of SSD I/O behavior (i.e., program, erase, and read operations) and its detailed FTL management policies, which include (1) L2P address translation, (2) request scheduling across channels and flash dies, (3) garbage collection and wear-leveling, (4) the NVMe interface over PCIe for host-device communication, and (5) contention on shared resources such as flash channels and dies.
We implement a demand-based L2P mapping cache (e.g., DFTL~\cite{gupta2009dftl}) based on MQSim's DRAM cache modeling. Only a subset of mapping entries is cached in SSD DRAM, and the remaining entries are fetched from flash chips on demand.

\head{NDP Extensions}
To enable realistic modeling of NDP in SSDs, we implement five key extensions in \namePaper{}'s simulator.  

First, we add a detailed internal DRAM architecture to our simulator based on \rncamv{Ramulator 2.0}~\cite{ramulator-github, ramulator-pim, luo2023ramulator2, kim-cal-2016} to capture bank-level parallelism, timing constraints (e.g., tRCD, tRP, tRAS), and DRAM channel bandwidth (see Table~\ref{tab:eval_config} for our simulated \rncamv{SSD-internal DRAM} characteristics). 
Second, we add the computation models of three NDP paradigms: \isp{}, \pud{}, and \ifp{}. We describe the NDP paradigms in \S\ref{subsec:background_ndp} and their performance modeling in \S\ref{subsec:methodology_modeling}.
Third, we add a dedicated execution queue to each SSD computation resource to accurately capture its utilization. 
Fourth, we extend MQSim's request scheduling mechanism to work in conjunction with \namePaper{}'s offloader. 
\namePaper{} uses this scheduling policy to move operands across computation resources and dispatch computation instructions to execution queues of computation resources.
Fifth, we extend MQSim's page allocation policy to enforce the data layout constraints of each NDP paradigm. We describe \namePaper{}'s data mapping and internal data movement \rncamv{policies} in \S\ref{subsec:system_modifications}.

We design the computation units, scheduling policy, and resource queues as modular components, which can be easily disabled for SSD-only simulation. 

\begin{table}[t]
  \centering
  \footnotesize
  \renewcommand{\arraystretch}{1.2}
  \resizebox{\columnwidth}{!}{
  \begin{tabular}{|c||l|}
  \hline
      
  \multirow{11}{*}{
  \shortstack{\textbf{SSD~\cite{mqsim-github, tavakkol2018mqsim}} \\ \\ \textbf{(Simulated)}}}
  & 48-WL-layer 3D TLC NAND flash-based SSD~\cite{samsung-980pro}; 2 TB\\ \cline{2-2}
  & \textbf{External Bandwidth}: PCIe 4.0, 8GB/s\\ \cline{2-2} 
  & \textbf{NAND Config}: 8 channels; 8 dies/channel; 2 planes/die;\\
  & 2,048 blocks/plane; 196 (4$\times$48) WLs/block; 4 KiB/page\\ \cline{2-2} 
  & \textbf{Flash Channel Bandwidth}: 1.2 GB/s\\ \cline{2-2} 
  & \textbf{Latency}: 
  T\textsubscript{read} (SLC mode): 22.5 $\mu$s~\cite{park2022flash}; \\
  & T\textsubscript{prog} (SLC mode): 400  $\mu$s~\cite{park2022flash}; T\textsubscript{bers} : 3500 $\mu$s~\cite{park2022flash}\\
  & T\textsubscript{AND/OR}: 20 ns~\cite{gao2021parabit}; T\textsubscript{latchtransfer}: 20 ns~\cite{gao2021parabit}; \\  
  &  T\textsubscript{XOR}: 30 ns~\cite{park2022flash}; T\textsubscript{DMA}: 3.3 $\mu$s;\\\cline{2-2} 
  & \textbf{Energy}: \texttt{E\textsubscript{read}} (SLC mode): 20.5$\mu$J/channel ~\cite{park2022flash}; \\ &
  E\textsubscript{AND/OR}: 10nJ/KB~\cite{gao2021parabit}; E\textsubscript{latchtransfer}: 10nJ/KB~\cite{gao2021parabit}; 
  \\ & E\textsubscript{XOR}: 20nJ/KB~\cite{park2022flash}; E\textsubscript{DMA}: 7.656$\mu$J/channel; \\
  \hline

  \multirow{4}{*}{\centering \shortstack{\textbf{SSD DRAM} \\ \textbf{(Simulated)}~\cite{ramulator-github, ramulator-pim, luo2023ramulator2, kim-cal-2016}}} 
  & \rncamv{2GB LPDDR4-1866~\cite{oh20143, jedec2021lpddr4} DRAM};\\
  & 1 channel, 1 rank, 8 banks\\ \cline{2-2} 
  & \textbf{Latency:} T\textsubscript{bbop}: 49 ns;  \textbf{Energy:} E\textsubscript{bbop}: 0.864 nJ; \\ 
  & where $bbop$ is bulk bitwise operation \\ 
  \hline 
  
  \multirow{2}{*}{\centering \shortstack{\textbf{SSD Controller Cores} \\ \textbf{(Emulated)~\cite{qemu}}}} 
  & \multirow{2}{*}{
    \rncamv{5 ARM Cortex-R8 cores @1.5 GHz~\cite{arm-cortexR8}}
    } \\
    & \\
  \hline

  \multirow{5}{*}{
  \centering 
  \shortstack{\textbf{Host CPU} \\ \textbf{(Real System)}}}
  & Intel(R) Xeon(R) Gold 5118\rncamv{~\cite{xeon5118}} \\ \cline{2-2} 
  & x86-64~\cite{guide2016intel}, 6~cores, out-of-order, 3.2 GHz  \\ \cline{2-2}
  & \emph{L1 Data + Inst. Private Cache:} 32kB, 8-way, 64B line \\ \cline{2-2}
  & \emph{L2 Private Cache:} 256kB, 4-way, 64B line \\ \cline{2-2}
  & \emph{L3 Shared Cache:} 8MB, 16-way, 64B line \\ 
  \hline

  \multirow{4}{*}{
  \centering 
  \shortstack{\textbf{Host GPU} \\ \textbf{(Real System)}}}
  & NVIDIA A100 GPU~\cite{a100, choquette2020nvidia}, \\
  & NVIDIA Ampere Architecture~\cite{ampere} (7nm) \\ \cline{2-2} 
  & 108 Streaming Multiprocessors, 1.4 GHz base clock;  \\ \cline{2-2}
  & \emph{L2 Cache:} 40 MB; Main Memory: 40GB HBM2~\cite{lee2016simultaneous, lee201425} \\ 
  \hline
  
  \multirow{2}{*}{\centering \shortstack{\textbf{Host Main Memory} \\ \textbf{\rncamv{(Simulated)}}}}
  & 32GB DDR4-2400\rncamv{~\cite{ddr4sheet, standard-jedec-2020}}, 4~channels, 1 rank, 16 banks; \\ \cline{2-2} 
  & Peak throughput: 19.2 GB/s; \rncamvi{FR-FCFS scheduling~\cite{mutlu2007stall, rixner2000memory, rixner2004memory, zuravleff1997controller}} \\ \cline{2-2} 
  \hline
  
  \end{tabular}
  }
\caption{Evaluated Configurations.}
\label{tab:eval_config}
\end{table}

\subsection{Performance and Energy Modeling \label{subsec:methodology_modeling}}

\noindent \textbf{Performance Modeling.} We model the performance of computations performed on three SSD computation resources. 
\squishlist 
\item \textbf{Computation Using Flash Memory Chips:} We model: (i) command transfer latency, 
(ii) flash sensing latency (i.e., $t_R$), 
(iii) data transfer from page buffer to flash controller (i.e., $t_{DMA}$),
(iv) program latency of Enhanced SLC-mode programming technique~\cite{park2022flash},
(iv) arithmetic operations (i.e., addition, multiplication) using the peripheral circuitry (e.g.,~\cite{chen2024aresflash, kabra2025ciphermatch}) (see \S\ref{sec:background}), 
(v) bitwise \rncamv{operations} via multi-wordline sensing (MWS)~\cite{park2022flash} (see \S\ref{sec:background}).
We rigorously validate our flash \rncamv{memory} model based on results from prior works~\cite{chen2024aresflash, kabra2025ciphermatch, park2022flash}.

\item \textbf{Computation Using DRAM:} We model (i) SSD DRAM access latencies, (ii) data movement from NAND flash memory to SSD DRAM using flash read and DMA operations, and (iii) bitwise and arithmetic operations based on SIMDRAM, MIMDRAM and Proteus frameworks~\cite{hajinazar2021simdram, oliveira2024mimdram, de2025proteus} (see \S\ref{sec:background}).

\item \textbf{Computation \rncamv{on} SSD Controller Cores:} We model computation latencies of the SSD controller cores using QEMU-based emulation~\cite{qemu}, configured for ARM Cortex R8~\cite{arm-cortexR8}.
\squishend

For runtime offloading, \namePaper{} tracks the: (i) latency of data movement between SSD computation resources, (ii) queuing delays of SSD controller core, NAND flash chips, and SSD DRAM, and (iii) computation latency of the offloaded instructions. We provide a detailed description of these components in \S\ref{subsec:mechanism_runtimeoffloading}.

\noindent\noindent\textbf{Energy Modeling.} 
We model the energy consumption of (1) computation on each SSD computation resource, and (2) data movement between the host and SSD, and across SSD computation resources.  
For NAND flash memory, we use the SSD power values of Samsung 980 Pro SSDs~\cite{samsung-980pro} and the NAND flash power values measured by Flash-Cosmos~\cite{park2022flash}. 
We model the energy consumption of computations on the controller cores using power models from prior works~\cite{cho2015design, harris2020ultra, mohan2013modeling, arm-cortexR8, qemu}.
For SSD DRAM, we use the power values of DDR4 DRAM\rncamv{~\cite{ghose2019demystifying, ghose2018your}}. 

\subsection{Evaluated Techniques \label{subsec:methodology_baseline}}

\head{Host CPU and GPU} 
We execute all computations on the host CPU and GPU (see Table~\ref{tab:eval_config} for our CPU and GPU configurations) by reading operands from the SSD via host memory. The operands are transferred from the SSD to the host memory through the NVMe interface.
\rncamv{We evaluate host CPU and GPU on a real system to establish strong baselines using architecture-optimized implementations (e.g., \rncamv{optimized} CUDA for GPU).
To strengthen the baselines, we combine real-system-based host CPU and GPU execution with simulation-based modeling of SSD-to-host data transfers. 
We assume full PCIe 4.0 bandwidth availability for the host baselines to prevent PCIe contention from degrading computation performance. 
We model the host \rncamvi{main memory (including DRAM latency and DRAM bandwidth)} using parameters in Table~\ref{tab:eval_config}.
}

\head{NDP Baselines} We evaluate four baseline NDP techniques and two prior offloading policies against \namePaper{}.   
First, \fc{}~\cite{park2022flash} proposes in-flash bulk bitwise operations. \fc{} enables bitwise AND on 48 operands located within the same block and bitwise OR on four operands located in different blocks within the same plane. 
Second, \af{}~\cite{chen2024aresflash} extends in-flash processing by supporting both bulk bitwise and arithmetic operations. 
Third, Processing-using SSD-DRAM (\pud{}) uses the MIMDRAM framework~\cite{oliveira2024mimdram} for computations on SSD DRAM. It leverages the detailed DRAM subsystem modeling in our simulator with computation latencies derived from DRAM row-activation and bus timing.  
\rncamv{Fourth,} \isp{} performs all computations on embedded general-purpose cores in the SSD controller. \fc{}, \af{}, and \pud{} leverage the SSD controller cores to execute computations that they do not support.  
Fifth, the bandwidth-based offloading model (\bw{})~\cite{hsieh.isca16, yang2023lambda, wolski2008using,  hadidi2017cairo, wu2020tuning, alsop2024pim} offloads each instruction to a computation resource that has the lowest bandwidth utilization among the SSD computation resources. 
Sixth, the data-movement-based offloading model (\dm{})~\cite{ghiasi2022alp, kim2017toward, wei2022pimprof,jiang20243} offloads each instruction to a computation resource that minimizes operand data movement.

\head{Ideal Approach} 
We implement an \ideal{} policy that assumes (1) no queueing delays in each computation resource, (2) zero data movement latency when operands are moved across resources, and (3) selection of a target computation resource with the least computation latency. This approach is \rncamv{\emph{not}} realizable in practice, but serves as an upper bound of performance when exploiting all SSD computation resources without runtime bottlenecks.

\begin{figure*}[b]
\centering
\includegraphics[width=\textwidth]{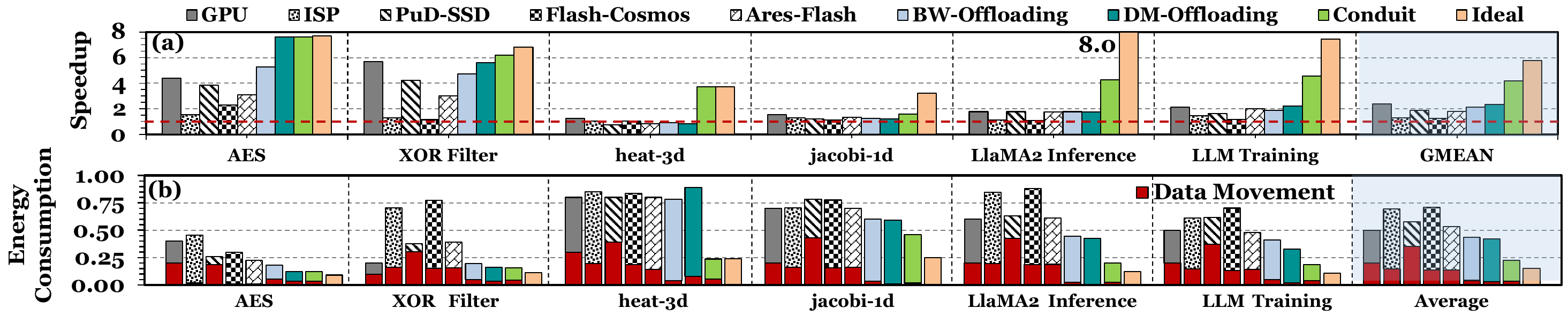}
\caption{Speedup (a) and energy consumption (b) of \namePaper{} and baselines normalized to those of \cpu{}. Each bar in Figure (b) shows energy consumption of data movement (in red) and computation (rest of the bar).}
\label{fig:Speedup}
\end{figure*}

\subsection{Evaluated Workloads \label{subsec:workloads}}
We evaluate \namePaper{} and baselines on six data-intensive workloads from diverse benchmark suites (e.g., polybench~\cite{pouchet2012polybench}, Rodinia~\cite{che2009rodinia}, LLMs~\cite{Karpathy}) (see Table~\ref{tab:workloads_characterization}). 
We select these workloads to capture diverse computation patterns, data movement characteristics, and vectorization coverage.
We characterize each workload based on three characteristics.
First, the vectorizable code percentage represents the fraction of application code that is automatically vectorized by \namePaper{}'s compile-time preprocessing (see \S\ref{subsec:mechanism_offline_preprocessing}). 
This defines the amount of computation suitable for fine-grained instruction-granularity offloading.
Second, average reuse measures the average number of operations that consume the same data before it is replaced or evicted. Average reuse directly affects data movement across SSD computation resources.
Third, we capture the mix of low-latency (e.g., bitwise, logical), medium-latency (e.g., add, predication), and high-latency (e.g., multiplication) operations in each workload. 
This influences the runtime selection of SSD computation resources.

We quantize floating-point operations to integer (INT8) because SSD computation resources lack native floating-point arithmetic support. This quantization enables the complete execution of all workloads within the SSD.
The memory footprint of each workload exceeds the SSD capacity by 2$\times$, inducing resource contention and data movement scenarios. 

Our workloads include \rncamv{the following}:

\noindent (1) \textbf{Advanced Encryption Standard (AES)}~\cite{noauthor_patmos_hlsbenchmarks_nodate, hara2008chstone, daemen1999aes, daemen2002data}. 256-bit encryption and decryption algorithm with high data reuse and low-latency bitwise operations, which makes it suitable for \ifp{}.

\noindent (2) \textbf{XOR Filter}~\cite{noauthor_hexopsfastfilter_2024, graf2020xor, graf2022binary}. A probabilistic data structure for fast membership tests similar to Bloom filters~\cite{bloom1970space}, dominated by arithmetic and predication operations, which execute efficiently in flash or DRAM.

\noindent (3) \textbf{heat-3d}~\cite{pouchet2012polybench}. A three-dimensional stencil computation that repeatedly updates each grid point using its neighbors. heat-3d has moderate-to-high arithmetic intensity with high data reuse across time steps, which benefits from coordinated offloading across multiple SSD computation resources. 

\noindent (4) \textbf{jacobi-1d}~\cite{pouchet2012polybench}. A one-dimensional stencil computation solver that updates each element using values from its immediate neighbors. jacobi-1d has moderate-to-high latency operations and data dependencies that benefit from offloading across multiple SSD computation resources.

\noindent (5) \textbf{LlaMA2 Inference}~\cite{Karpathy, touvron2023llama}. INT8 inference of \rncamv{the 7B-parameter} LLaMA2 model~\cite{touvron2023llama} with a mix of high- and medium-latency operations.

\noindent (6) \textbf{LLM Training}~\cite{Karpathy, touvron2023llama}. Bandwidth-intensive INT8 training of \rncamv{the 7B-parameter} LLaMA2 model~\cite{touvron2023llama}, characterized by frequent weight updates, data movement, and a combination of arithmetic and control-intensive operations.

\begin{table}[t]
    \centering
    \setlength{\tabcolsep}{2pt}
    \scriptsize
    \resizebox{\linewidth}{!}{
    	\begin{tabular}{|c||c|c|c|c|c|}\hline
    	\begin{tabular}{c}\textbf{Workload} \end{tabular} &	
        \begin{tabular}{c}\textbf{Vectorizable} \\ \textbf{Code\%} \end{tabular} 
        & \textbf{Avg. Reuse} & 
            \begin{tabular}{c}    
            \textbf{Low} \\\textbf{Latency} \\\textbf{Operations}
            \end{tabular} &
            \begin{tabular}{c}    
            \textbf{Medium} \\\textbf{Latency} \\\textbf{Operations}
            \end{tabular}
             &
             \begin{tabular}{c}    
            \textbf{High} \\\textbf{Latency} \\\textbf{Operations}
            \end{tabular} \\
            \hline
            \hline
                            
            \begin{tabular}{c}\textbf{AES}\\~\cite{noauthor_patmos_hlsbenchmarks_nodate, hara2008chstone, daemen1999aes, daemen2002data}
            \end{tabular}
            & 65\% & 15.2 & 87\% & 13\% & 0\% \\\hline

            \begin{tabular}{c}\textbf{XOR-Filter}\\~\cite{noauthor_hexopsfastfilter_2024, graf2020xor, graf2022binary}
            \end{tabular}
            & 16\% & 2.0 & 1\% & 98\% & 1\% \\\hline

            \begin{tabular}{c}\textbf{heat-3d}~\cite{pouchet2012polybench}
            \end{tabular}
            & 95\% & 16 & 0\% & 60\% & 40\% \\\hline

            \begin{tabular}{c}\textbf{jacobi-1d~\cite{pouchet2012polybench}}
            \end{tabular}
            & 95\% & 3 & 0\% & 67\% & 33\% \\\hline

            \begin{tabular}{c}\textbf{LlaMA2} \\ \textbf{Inference}~\cite{Karpathy, touvron2023llama}
            \end{tabular}
            & 70\% & 1.8 & 0\% & 53\% & 47\% \\\hline

            \begin{tabular}{c}\textbf{LLM} \\ \textbf{Training}~\cite{Karpathy, touvron2023llama}
            \end{tabular}
            & 60\% & 5.2 & 0\% & 88\% & 12\% \\\hline
    	\end{tabular} 
	}
    \caption{Characteristics of the Evaluated Workloads.} 
    \label{tab:workloads_characterization}
\end{table}

\section{Evaluation \label{sec:evaluation}}

\subsection{Performance Analysis \label{subsec:perf_analysis}}

\fig~\ref{fig:Speedup}(a) shows the speedup of \namePaper{} and the baselines normalized to that of \cpu{} across six real-world applications. We make four key observations.

First, \namePaper{} outperforms all baselines, achieving average speedup of $4.2\times$ over CPU, $1.8\times$ over GPU, $3.3\times$ over \isp{}, $2.2\times$ over \pud{}, $3.3\times$ over \fc{}, $2.3\times$ over \af{}, $2.0\times$ over \bw{} and $1.8\times$ over \dm{}.

Second, \namePaper{} outperforms the best-performing prior offloading model, \dm{}, by $1.8\times$ on average by making holistic offloading decisions that account for factors such as resource queueing delays and data dependence delays, which \dm{} does not consider. By prioritizing data movement reduction, \dm{} frequently offloads instructions to the same computation resource (e.g., flash chips), which leads to resource contention and significant queueing delays.

Third, \namePaper{} \rncamvi{provides} $62\%$ of \rncamvi{the} \ideal{} offloading approach's performance, which demonstrates \namePaper{}'s ability to effectively exploit SSD computation resources without the ideal approach's \rncamvi{unrealistic} assumptions. 

Fourth, \namePaper{} provides higher performance benefits  ($2.6\times$ on average over \dm{}, up to $4.5\times$)  over prior offloading approaches for compute-intensive workloads (e.g., heat-3d, jacobi-1d, LlaMA2 Inference, LLM Training). These workloads exhibit high vectorization coverage and a large fraction of medium- and high-latency operations (see Table~\ref{tab:workloads_characterization}), which makes them sensitive to both computation throughput and resource contention.
\namePaper{} dynamically exploits multiple computation resources effectively, whereas prior offloading approaches prefer specific resources (e.g., \ifp{} for \dm{}).
For memory-bound workloads (e.g., AES, XOR Filter), prior offloading approaches achieve performance comparable (\namePaper{} outperforms \dm{} by $1.2\times$ on average) to \namePaper{} because data movement is a key factor in resource selection.

We conclude that \namePaper{} provides significant performance gains by dynamically \rncamvi{and judiciously} exploiting multiple computation resources. 

\subsection{Energy Analysis \label{subsec:energy_analysis}}
\fig~\ref{fig:Speedup}(b) shows the energy consumption of \namePaper{} and the baselines normalized to that of CPU.  Each bar in \fig~\ref{fig:Speedup}(b) shows a breakdown of data movement energy (in red) and computation energy (the rest of the bar). We make five key observations.

First, \namePaper{} has lower average energy consumption than all baselines, reducing energy consumption by $78.2\%$ over CPU,  $58.2\%$ over GPU,  $67.3\%$ over \isp{},  $60.6\%$ over \pud{},  $68.0\%$ over \fc{},  $57.4\%$ over \af{},  $47.8\%$ over \bw{}, and  $46.8\%$ over \dm{}. \namePaper{} reduces data movement and the overall execution time by efficiently exploiting multiple SSD computation resources. 
Second, \namePaper{} \rncamvi{provides} $68\%$ of the energy efficiency of \rncamvi{the} \ideal{} approach, which assumes no resource contention.
Third, \dm{} is the most energy-efficient prior offloading approach because it explicitly reduces data movement across computation resources. \namePaper{} reduces energy consumption by $46.8\%$ on average over \dm{} by jointly considering resource contention with data movement costs. 
Fourth, \namePaper{} reduces energy consumption by $57\%$ over \af{}, the most energy-efficient single-resource NDP technique. While \af{} has lower data movement energy, it relies on \isp{} to execute computations that it does not support.
Fifth, although GPU offers high computation throughput, it incurs $58.2\%$ more energy than \namePaper{} due to significant data movement over PCIe and high computation power. 

We conclude that \namePaper{} provides significant energy benefits by exploiting SSD-internal parallelism, reducing data movement, and avoiding expensive host-side computation.

\subsection{Tail Latency \label{subsec:tail_latency}}

\rncamvi{\figs{~\ref{fig:tail_latency}(a)}} and \ref{fig:tail_latency}(b) show the 99th and 99.99th percentile I/O latencies (\emph{tail latencies}) of \ideal{}, \namePaper{}, \bw{}  and \dm{} using two representative workloads, LlaMA2 Inference, and jacobi-1d (see \S\ref{subsec:workloads}).

\namePaper{} significantly reduces tail latency compared to the baselines across the two workloads. 
\rncamvi{In LlaMA2 Inference, compared to \bw{} (\dm{}), \namePaper{} reduces the 99th percentile latency by $1.8\times$ ($5.6\times$), and 99.99th percentile latency by $10.7\times$ ($22.3\times$).
In jacobi-1d, compared to \bw{} (\dm{}), \namePaper{} reduces the 99th percentile latency by $1.7\times$ ($1.1\times$), and 99.99th percentile latency by $1.9\times$ ($1.3\times$).}
\namePaper{}'s tail latency \rncamvii{benefits stem} from its resource contention-aware offloading, which balances execution across multiple computation resources.

\begin{figure}[t]
\centering
\includegraphics[width=0.9\linewidth]{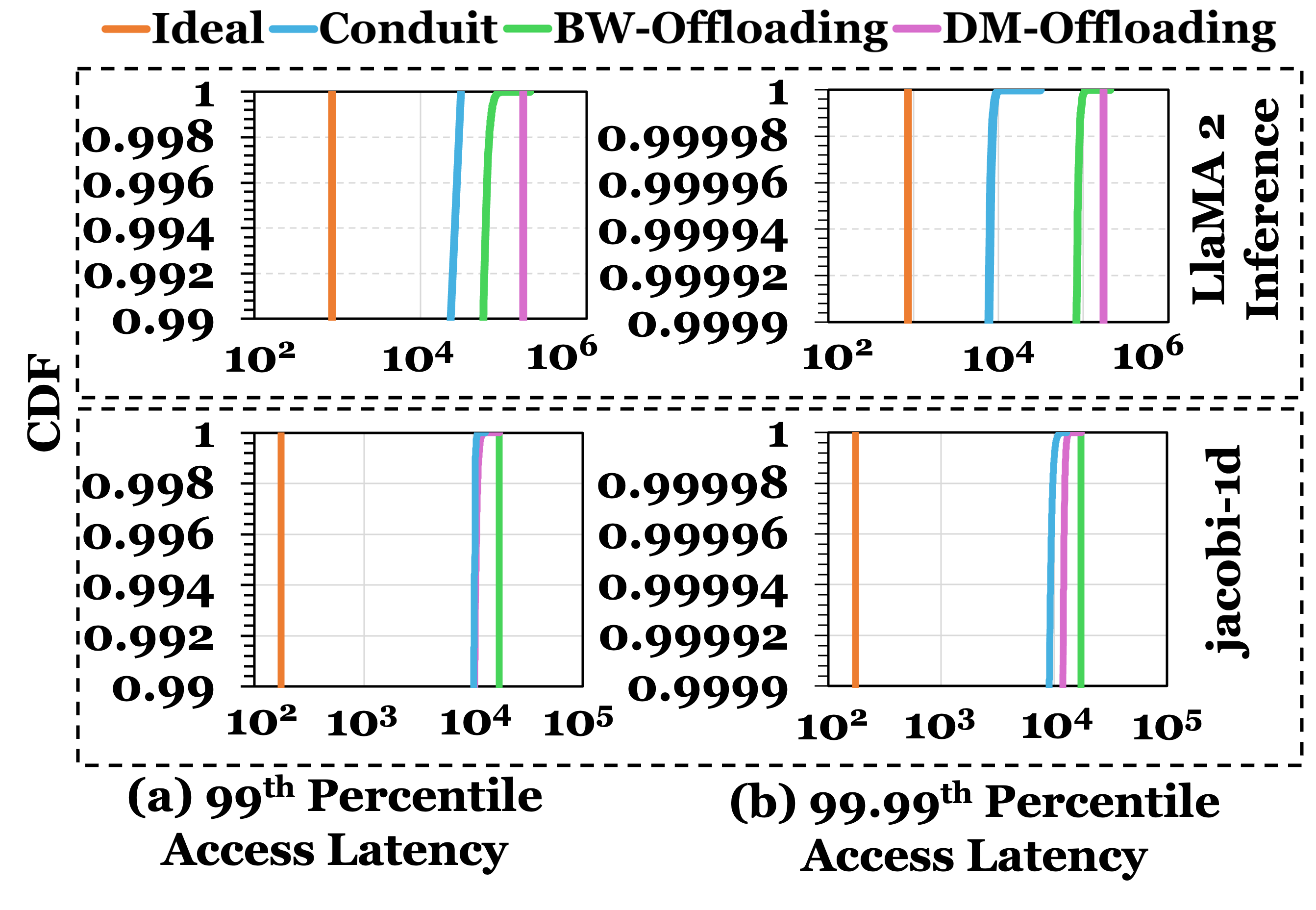}
\caption{Comparison of (a) 99th percentile and (b) 99.99th percentile of I/O request latencies of \ideal{}, \namePaper{}, \bw{}, \dm{} in LlaMA2 Inference and jacobi-1d.}
\label{fig:tail_latency}
\end{figure}

\subsection{Offloading Decisions}
\fig~\ref{fig:Decisions} shows the offloading decisions of Ideal, \namePaper{}, \dm{} and \bw{}. \rncamvi{Offloading} decisions are represented as the fraction of instructions offloaded to each computation resource. We make three key observations.

\begin{figure}[b]
\centering
\includegraphics[width=\linewidth]{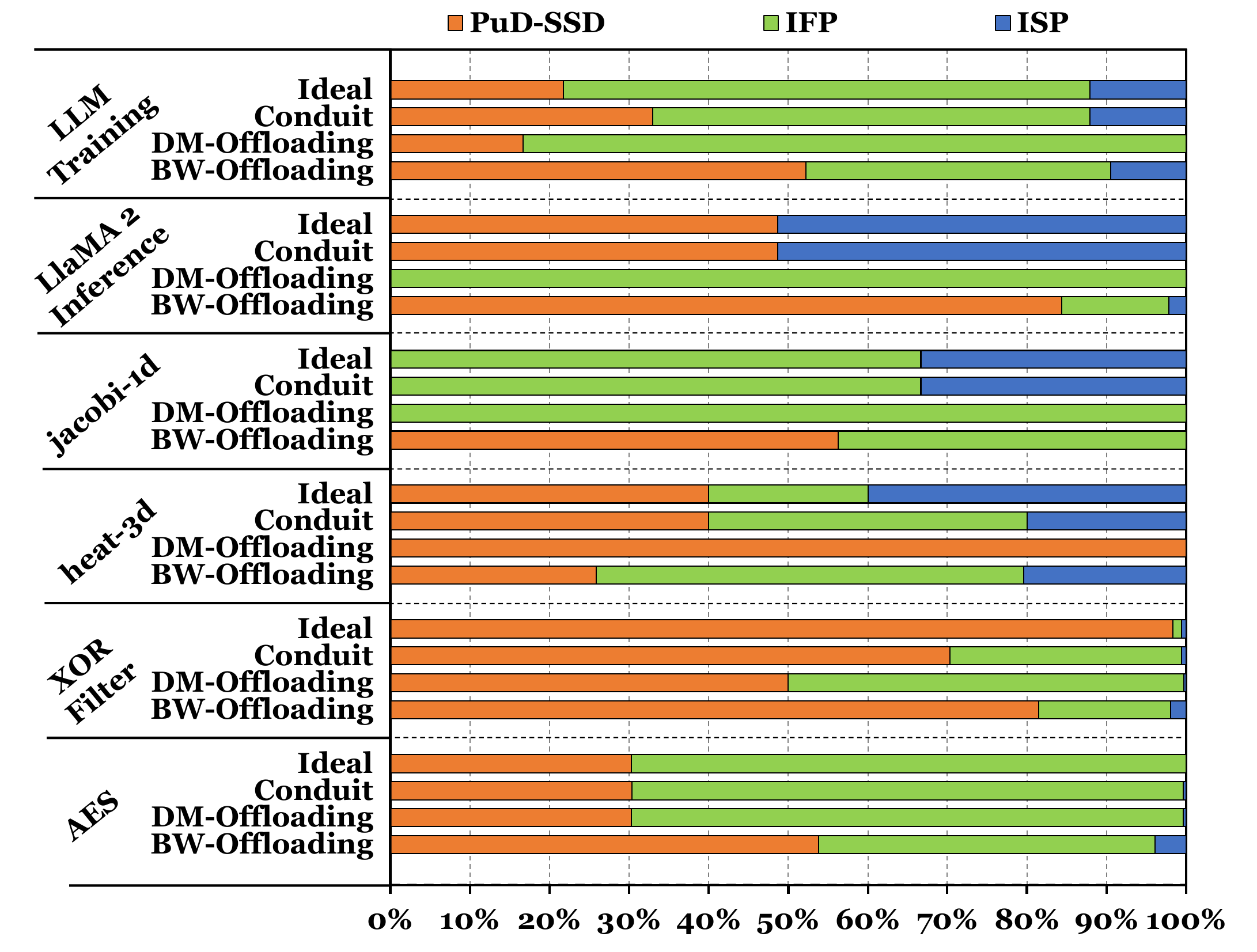}
\caption{Fraction of Instructions Offloaded to Each Computation Resource in \ideal{}, \namePaper{}, \dm{} and \bw{}.}
\label{fig:Decisions}
\end{figure}

First, \namePaper{}'s computation resource utilization closely matches that of an \ideal{} policy across most workloads. 
Second, in memory-bound workloads (e.g., AES, XOR-Filter), all approaches, including \namePaper{}, utilize \isp{} very sparingly (\namePaper{} offloads $0.4\%$ of instructions to \isp{} in AES and $0.6\%$ in XOR Filter). \isp{}'s limited SIMD parallelism and significant data movement from the flash chips to the SSD controller make \ifp{} and \pud{} more efficient \rncamvi{(see \S\ref{subsec:motivation_casestudy})}. 
Third, in compute-intensive applications (e.g., LLM Training, heat-3d), \namePaper{} and \ideal{} distribute computations across multiple resources. In LlaMA 2 Inference, both \namePaper{} and \ideal{} split execution almost equally between \pud{} and \isp{}. Both policies avoid \ifp{} because multiplication operations in \ifp{} require frequent operand transfers between the flash controller and flash chips for shift-and-add \rncamvi{operations}~\cite{chen2024aresflash}.

\rncamvi{We conclude that \namePaper{} successfully selects the \rncamvii{resource to use for computation} based on characteristics of the executing workload.}

\subsection{Workload-Computation Resource Interaction \label{subsec:results_interaction}}
To analyze the interaction between the workload and computation resources, we observe the operations and the computation resources chosen by different offloading policies during the execution of LlaMA2 Inference~\cite{Karpathy}. 
\fig~\ref{fig:workload_computation_interaction}(a) shows the different operations (i.e., addition, subtraction, multiplication, and shuffle) in the workload.
\figs~\ref{fig:workload_computation_interaction}(b), \ref{fig:workload_computation_interaction}(c), \ref{fig:workload_computation_interaction}(d) show the computation resources chosen by \bw{}, \dm{} and \namePaper{} respectively. 
We show the execution of 12000 vectorized instructions.

\begin{figure}[b]
\centering
\includegraphics[width=\linewidth]{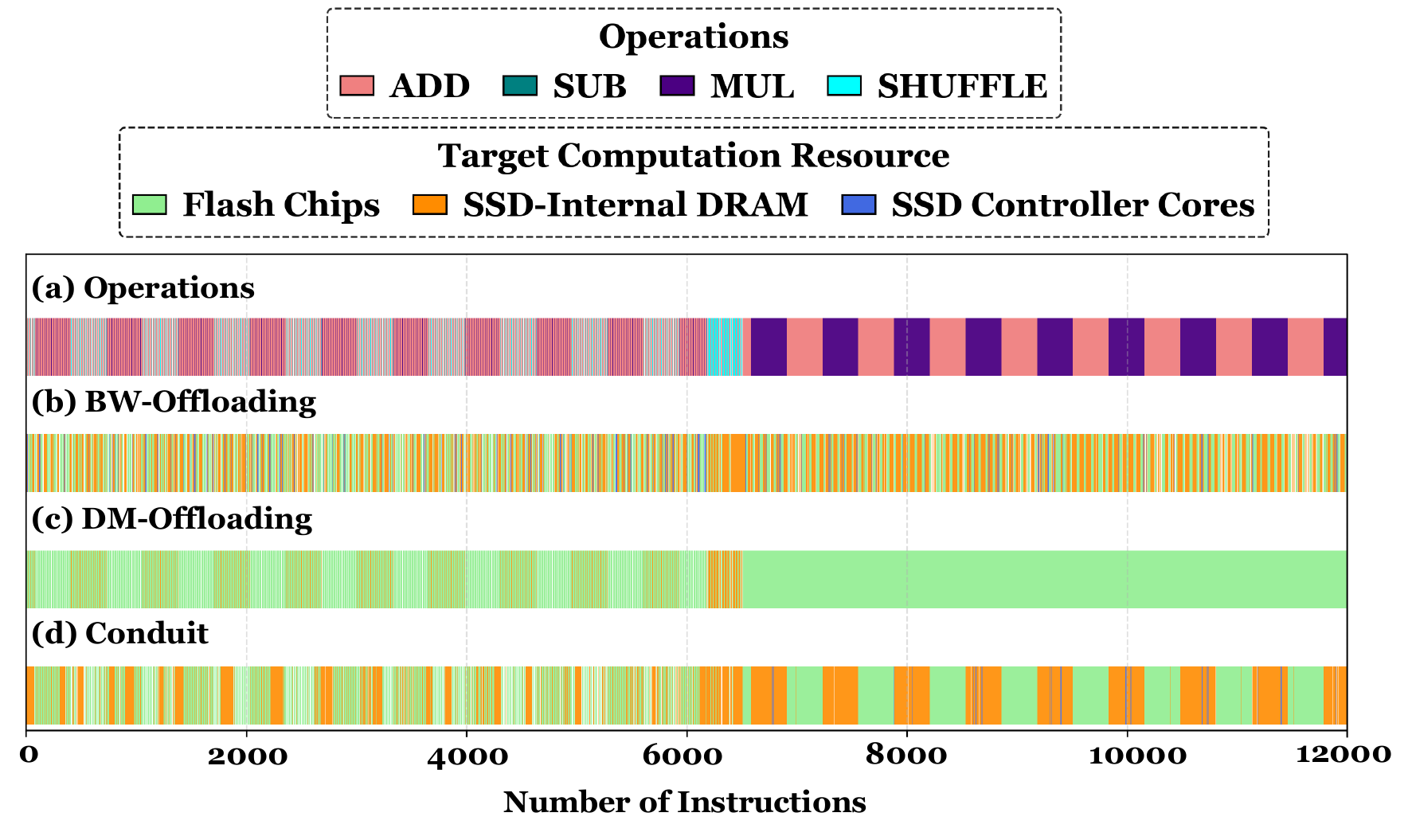}
\caption{Instructions to computation resource mapping in \namePaper{} and prior offloading policies.}
\label{fig:workload_computation_interaction}
\end{figure}

We make three key observations.
First, \bw{} frequently switches between flash, DRAM, and controller cores to balance bandwidth, which results in frequent data transfers.
Second, \dm{} executes addition and multiplication operations (instructions 6000-12000) in flash to minimize data movement, but this leads to flash contention and underutilization of other SSD computational resources.
Third, \namePaper{} dynamically adapts resource selection based on instruction type and runtime conditions.
It executes locality-friendly addition operations (i.e., pink phases after 6000 instructions) in flash while performing costly multiplication operations (i.e., purple phases after 6000 instructions) in DRAM and control-intensive operations in the controller cores. Unlike \bw{}, \namePaper{} balances computation across computation resources less aggressively, which reduces synchronization and data movement overheads.
Overall, Conduit \rncamvi{adapts its decisions to both workload phases and resource availability} and \rncamvi{thereby} delivers higher performance than prior offloading policies.

\section{Discussion \label{sec:discussion}}
\head{\namePaper{}'s Extensibility} 
\namePaper{} leverages existing commodity SSD computation resources, but it can be extended to support new operations \rncamvi{(e.g., sort~\cite{vermij2017sorting, stonebraker2013voltdb, soysal2025mars, mirzadeh2015sort, quero2015self, lee2016activesort, salamat2021nascent, myung2020efficient, salamat2022nascent2, lee2015external, wu2015data, wang2016ssd}, search~\cite{kabra2025ciphermatch, chen2024search, wang2024ndsearch, bundy1984breadth, altschul-jmb-1990, zhang2021max, wang2016ssd1})} or specialized hardware accelerators \rncamvi{(e.g.,~\cite{ghiasi2022genstore, ghiasi_megis_2024, chen2025reis, jun2015bluedbm, matam2019graphssd, wang2024beacongnn, li2023ecssd, kim-fast-2021, wang2024ndsearch, hu2022ice, jun-hpec-2016, pan2024instattention})}. 
\namePaper{}'s extensibility to future hardware capabilities and application demands requires adding new application and hardware-specific characteristics in its cost function.

\head{Limitations of Auto-Vectorization} 
The compiler's auto-vectorization technique can fail for application code regions with 
(1) \rncamvi{complex} data dependencies, memory aliasing, or indirect accesses, 
(2) complex control flow or multiple exit points, 
(3) atomic or synchronized operations, and 
(4) loops with unknown or small iteration counts. 
As a result, \namePaper{} may not fully leverage SIMD performance benefits. 
To mitigate this limitation, programmers can restructure the code, add compiler hints (pragmas), or manually vectorize.

\head{Applicability to Irregular Workloads}
\namePaper{} is effective not only for workloads dominated by computation-heavy loops, but also for irregular workloads (e.g., database sorting, merging, aggregation) that contain both computation-heavy loops and control-intensive code regions (e.g., branches and conditional statements)\rncamvi{, as we show in \S\ref{subsec:workloads}}. Such workloads are challenging for single-resource NDP because different workload phases exhibit different execution characteristics.

\namePaper{} address this challenge by leveraging multiple computation resources with diverse computation capabilities. \namePaper{} offloads vectorizable, data-parallel code sections to \pud{} and \ifp{}, which exploit SIMD parallelism. It executes control-intensive code regions on general-purpose SSD cores, which provide flexible control flow and low data-movement overhead for non-vectorizable code. By dynamically matching each code region to the most suitable computation resource, \namePaper{} efficiently supports a broad class of applications.

\section{Related Work \label{sec:related_work}}
To our knowledge, \namePaper{} is the first general-purpose programmer-transparent NDP framework that dynamically exploits multiple heterogeneous SSD computation resources. 
We \rncamvi{already qualitatively and quantitatively compared} \namePaper{} against (1) state-of-the-art NDP approaches, \pud{} (e.g.,\cite{hajinazar2021simdram, oliveira2024mimdram}), Flash-Cosmos~\cite{park2022flash} and \af{}~\cite{chen2024aresflash}, and (2) two NDP offloading models from other domains, \bw{} (e.g.,~\cite{hsieh.isca16, yang2023lambda, wolski2008using,  hadidi2017cairo, wu2020tuning, alsop2024pim}) and \dm{} (e.g.,~\cite{ghiasi2022alp, kim2017toward, wei2022pimprof,jiang20243}). 
We review related works on SSD-based NDP techniques and NDP offloading.

\head{Application-Specific NDP Using Multiple SSD Computation Resources} 
\rncamvi{Various} works \rncamvi{(e.g.,~\cite{jang2025inf,pan2024instattention, chen2025reis, soysal2025mars, wang2024beacongnn, yu2024cambricon, chen2024search, kabra2025ciphermatch, wang2024ndsearch, hu2022ice, li2023ecssd, wang2016ssd, wang2016ssd1, lincoln-hpca, lee2025aif, liang-fpl-2019, park2016storage, mailthody-micro-2019, kim-fast-2021, kang-msst-2013, torabzadehkashi-pdp-2019, wang-eurosys-2019,acharya-asplos-1998,keeton-sigmod-1998, tiwari-fast-2013, tiwari-hotpower-2012, boboila-msst-2012, bae-cikm-2013,torabzadehkashi-ipdpsw-2018, pei-tos-2019, do-sigmod-2013, kim-infosci-2016, liang-atc-2019,cho-wondp-2013, jun2015bluedbm, lee-ieeecal-2020, ajdari-hpca-2019, jun-hpec-2016, kang-tc-2021, kim-sigops-2020, lee2022smartsage, ruan2019insider, li-atc-2021, jeong-tpds-2019, mao2012cache, gouk2024dockerssd, ghiasi_megis_2024, ghiasi2022genstore, kang-micro-2021, yavits2021giraf, kim2023optimstore, lim-icce-2021, narasimhamurthy2019sage, jun-isca-2018, fakhry2023review, gu-isca-2016, jo2016yoursql, mahapatra2025rag, lee2017extrav,  pan2024instinfer})} propose domain-specific NDP designs that leverage multiple SSD computation resources for targeted applications such as genomics, nearest neighbor search, retrieval-augmented generation, \rncamvi{graph processing,} and LLM training/inference acceleration.
MARS~\cite{soysal2025mars} accelerates raw signal genome analysis~\cite{firtina2023rawhash, firtina2024rawhash2, firtina2024rawsamble, lindegger2024rawalign, eris2025rawbench} by performing computations within the SSD-internal DRAM and the SSD controller. 
REIS~\cite{chen2025reis} accelerates approximate nearest neighbor search (ANNS) kernels by offloading computations to the peripheral circuitry of flash chips and the SSD controller.
NDSearch~\cite{wang2024ndsearch} offloads graph traversal to the SSD controller and distance computations to hardware accelerators near flash chips.
CIPHERMATCH~\cite{kabra2025ciphermatch} accelerates homomorphic encryption-based string matching using in-flash arithmetic and comparison operations in the SSD controller. 
In contrast \rncamvi{to these specialized approaches}, \namePaper{} provides a general, application-transparent NDP framework that dynamically distributes computations across multiple heterogeneous SSD resources.

\head{NDP Offloading Approaches}
Several PIM frameworks identify offloadable code regions using profiling-based techniques \rncamvi{(e.g.,~\cite{wei2022pimprof, oliveira2021damov})} or compiler-based techniques \rncamvi{(e.g.,~\cite{yan2001codemap, nai2017graphpim, jaliminchecs, maity2024coat,chen2022offload, ahmed2019compiler, hadidi2017cairo, hsieh.isca16, ghiasi2022alp, peng_chopper_2023, devic2022pim, singh2025scalable, khadem2023vector})}. 
Prior works~\cite{maity2024coat,chen2022offload, ahmed2019compiler, hadidi2017cairo} propose PIM offloading techniques that incorporate data locality into their offloading strategies. 
Other works~\cite{kang2024isp,park2025selective,maity2025offload,olivier2019hexo} propose offloading frameworks that leverage a single computation resource near memory or storage. 
ISP Agent~\cite{kang2024isp} and SODE~\cite{park2025selective} propose changes across the storage stack to enable offloading to \isp{} in an SSD.
CORD~\cite{zaman2025cord} proposes techniques to offload query execution across multiple computational storage devices (CSDs) in a distributed environment.
Unlike these approaches, \namePaper{} proposes a general-purpose programmer-transparent offloading framework that dynamically leverages multiple SSD computation resources.

\section{Conclusion}
We introduce \namePaper{}, the first general-purpose programmer-transparent NDP framework that enables fine-grained instruction-granularity offloading across multiple heterogeneous SSD computation resources \rncamvi{(including SSD controller cores, SSD-internal DRAM\rncamvii{,} and flash chips)}.
\namePaper{} performs compile-time vectorization to identify offloadable code regions (e.g., loops with computations) and transforms them into SIMD operations. 
At runtime, \namePaper{} (i) determines the most suitable SSD computation resource using a holistic cost function, (ii) translates each vector operation to the native ISA of the chosen computation resource, and (iii) dispatches the transformed instruction to the chosen resource’s execution queue.
Our evaluation using an in-house simulator \rncamvi{(guided by real system execution)} on six real-world applications shows \namePaper{} outperforms the best-performing prior offloading policy by 1.8$\times$ and reduces energy consumption by 46\% on average.

\section*{Acknowledgments}
{
We thank the anonymous reviewers of MICRO 2025 and HPCA 2026 for their feedback. We thank the SAFARI Research Group members for providing a stimulating, inclusive, intellectual, and scientific environment.
We acknowledge the generous gifts from our industrial partners, including Google, Huawei, Intel, Microsoft, and VMware.
\rncamvi{This research was partially supported by European Union’s Horizon Programme for research and innovation under Grant Agreement No. 101047160 (project BioPIM), Swiss National Science Foundation (SNSF), Semiconductor Research Corporation (SRC), ETH Future Computing Laboratory (EFCL), Huawei ZRC Storage Team, and the AI Chip Center for Emerging Smart Systems Limited (ACCESS).}
}

\bibliographystyle{unsrt}


\end{document}